\documentclass[3p,preprint,12pt]{elsarticle}

\usepackage{amssymb}
\usepackage{algorithm}
\usepackage{algpseudocode}
\usepackage{amsmath}
\usepackage{xcolor}

\journal{Information Fusion}

\begin{document}

\begin{frontmatter}



\title{Comprehensive Botnet Detection by Mitigating Adversarial Attacks, Navigating the Subtleties of Perturbation Distances and Fortifying Predictions with Conformal Layers}

\author[label1]{Rahul Yumlembam}
\author[label1]{Biju Issac\corref{cor1}}
\ead{bissac@ieee.org}
\author[label2]{Seibu Mary Jacob}
\author[label1]{Longzhi Yang}
\affiliation[label1]{organization={Department of Computer and Information Sciences, Northumbria University},
            city={Newcastle upon Tyne},
            postcode={NE1 8ST},
            country={England, UK}}

\affiliation[label2]{organization={School of Computing, Engineering \& Digital Technologies, Teesside University},
            city={Middlesbrough},
            postcode={TS1 3BX},
            country={England, UK}}
\cortext[cor1]{Corresponding author}


\begin{abstract}
Botnets are computer networks controlled by malicious actors that present significant cybersecurity challenges. They autonomously infect, propagate, and coordinate to conduct cybercrimes, necessitating robust detection methods. This research addresses the sophisticated adversarial manipulations posed by attackers, aiming to undermine machine learning-based botnet detection systems. We introduce a flow-based detection approach, leveraging machine learning and deep learning algorithms trained on the ISCX and ISOT datasets. The detection algorithms are optimized using the Genetic Algorithm (GA) and Particle Swarm Optimization (PSO) to obtain a baseline detection method. The Carlini \& Wagner (C\&W) attack and Generative Adversarial Network (GAN)  generate deceptive data with subtle perturbations, targeting each feature used for classification while preserving their semantic and syntactic relationships, which ensures that the adversarial samples retain meaningfulness and realism. An in-depth analysis of the required L2 distance from the original sample for the malware sample to misclassify is performed across various iteration checkpoints, showing different levels of misclassification at different L2 distances of the Pertrub sample from the original sample. Our work delves into the vulnerability of various models, examining the transferability of adversarial examples from a Neural Network surrogate model to Tree-based algorithms. Subsequently, models that initially misclassified the perturbed samples are retrained, enhancing their resilience and detection capabilities. In the final phase, a conformal prediction layer is integrated, significantly rejecting incorrect predictions — 58.20\% in the ISCX dataset and 98.94\% in the ISOT dataset.

\end{abstract}




\end{frontmatter}


\section{Introduction}
\label{Introduction}
Cybercriminals can infect an organization's computing device or machine with a 'bot binary' executable using traditional attack techniques such as viruses and worms distributed through user downloads and email links. The bot binary runs silently in the background of a user machine, turning it into a 'zombie' waiting for commands from a Command and Control (C\&C) server controlled by the botmaster or another bot. A botmaster controls a group of zombies and forms a botnet to perform distributed computing tasks \cite{Thanh}. Cybercriminals have shifted in favour of botnet usage, with a recent report recording over 10,000 C\&C servers were added to the blocking lists \cite{Spamhaus}. For instance, in 2016, a DDoS attack carried out on DNS provider Dyn using the Mirai Internet of Things (IoT) Botnet caused users of the DNS service to have issues resolving domain names, ultimately causing many well-known sites to become inaccessible \cite{Dynstatus}. Cybercriminals frequently resort to botnets to earn profits, mainly through Cybercrime-as-a-Service (CaaS), where they rent out parts of a botnet to clients \cite{Putman}. Botmasters must ensure that their bot evades the Intrusion Detection Systems (IDSs) to maintain operational persistence, allowing them to gain buyers' trust. Doing so increases the potential for botnet growth, attracting higher profits from buyers requiring a large amount of cumulative bandwidth and processing power. 

Historically, the first line of defence against such threats has been Network Intrusion Detection Systems (NIDS). Traditional signature-based NIDS has revealed shortcomings, particularly in identifying novel or modified botnets. Consequently, a recent shift has been towards leveraging machine learning algorithms within NIDS. Machine learning systems promise to detect known threats and unearth previously unknown or zero-day threats. However, as with most advancements, this shift has vulnerabilities. Recent research underscores the susceptibility of machine learning models to adversarial attacks, especially evasion attacks. These manipulations create deceptive instances that could pass undetected, presenting a significant challenge.
While there has been extensive research on adversarial attacks in fields like computer vision, the implications for NIDS still need to be explored more in detail. The features used for training the machine learning-based NIDS system are central to its effectiveness. Therefore, understanding and assessing the vulnerabilities that arise when adversaries manipulate these features is paramount. However, the extent to which adversaries can exploit these features and the magnitude of feature-based perturbations needed to compromise NIDS remains unexplored and, therefore, raises pressing questions: How susceptible are these features to adversarial alterations? Can subtle feature manipulations successfully bypass a sophisticated machine-learning detector? In this research, we delve into these questions, aiming to discover the vulnerabilities of each feature used in the training. 
A common approach to generating adversarial examples involves using a surrogate model. When we choose a neural network as the surrogate model, an essential question arises: Will these adversarial examples maintain their deceptive potency against different architectures? In this research, we aim to find out the extent of transferability of the adversarial samples. 

First, we train both our machine learning and deep learning algorithms. To ensure the optimal performance of these models, we employ Genetic Algorithms (GA) and Particle Swarm Optimisation (PSO) to fine-tune the hyperparameters meticulously. This process not only aids in establishing a solid foundation but also in achieving the best predictive accuracy possible for our models.

In the next step, we focus on the individual features we use to train the machine learning model. To alter these features, we utilize the Carlini \& Wagner (C\&W) and Generative Adversarial Network (GAN) attacks. However, we maintain semantic and syntactic relationships during this attack by adjusting dependent features concerning the primary attack feature. By doing this, we seek to identify and understand potential vulnerabilities that might emerge when these features undergo manipulation. Moreover, having crafted these adversarial examples with a neural network as the surrogate model, we probe their efficacy against other machine learning architectures. Specifically, we test the transferability of these adversarial samples to different architectures, such as decision trees and random forests. This step is pivotal, as it helps us gauge the breadth of the threat, determining whether adversarial samples designed for one model can compromise another. Upon identifying potential vulnerabilities, we collect the adversarial samples. We then integrate these samples into our training process, further refining our models and enhancing their resilience against adversarial inputs.

Finally, this work introduces conformal prediction grounded in robust statistical principles. It offers a framework that makes predictions and provides a valid measure of certainty for each prediction. We can reject uncertain instances and accept only the classifications of confident ones, thereby offering an opportunity to investigate the rejected instances instead of making an incorrect prediction. The need for reliable prediction is paramount in domains such as NIDS. However, with evolving cyber threats and sophisticated adversarial tactics, traditional predictive models occasionally grapple with instances of uncertainty. The repercussions of false positives or negatives in a high-stakes environment like cybersecurity can be dire. Hence, there is an urgent need for methods that predict and quantify the confidence in these predictions.

In summary, the regular and novel contributions of this paper are as follows:
\begin{itemize}
    \item Optimised Model Training: We have taken a standard yet Discriminatoral approach in employing Genetic Algorithms (GA) and Particle Swarm Optimisation (PSO) to fine-tune the hyperparameters of our machine learning and deep learning models, which ensures we achieve competitive predictive accuracy with other works.
    \item In-depth Feature Vulnerability Analysis: Our research thoroughly examines the individual features utilized in training the machine learning model using C\&W and GAN attacks. We meticulously adjust dependent features with the primary attack feature to uncover potential vulnerabilities in the training features, making a novel contribution to feature analysis within machine learning models.
    \item Transferability Examination: In a novel exploration, we extend our work's impact by investigating how adversarial examples, designed using neural networks, might perform against different architectural paradigms, such as decision trees and random forests. This inquiry provides new insights and an understanding of model robustness and transferability.
    \item Introduction of Conformal Prediction in NIDS: We make a novel contribution by integrating and analyzing conformal prediction within Network Intrusion Detection Systems (NIDS). Utilizing this robust statistical principle-based approach, we offer a predictive framework that provides valid measures of prediction certainty and can reject uncertain instances.
\end{itemize}

\section{Related Ideas and Work}
\label{sec:related_works}

Botnets often communicate with Command and Control (C\&C) servers or other bots through network interactions. In order to identify such communications, Network Intrusion Detection Systems (NIDS) like Snort are deployed to identify them. These systems use signature-based detection to check each network packet against predefined signatures. A signature-based detection system matches predefined signatures to each packet's signature and generates an alert if a match is found, deeming it as a malicious behaviour \cite{Baker}. However, this requires analysis of every network packet, which is computationally intensive \cite{Roesch} and shown in research to suffer from a large proportion of packet drops when saturated at higher network speeds \cite{Shah}. Furthermore, experimentation shows that Snort's false positive rate (FPR) can be high, with the default rule sets, rendering it challenging to analyze or trust the alerts \cite{Shah}.
On the other hand, as these detection methods mature, bot developers innovate, too, crafting ways to dodge detection and ensure their bots remain stealthily active. Techniques include encrypting communication payloads and fragmenting packets. In sophisticated techniques like the polymorphic blending attack, a bot first understands the typical traffic profile of a network, it then mimics this profile when communicating, making activities blend seamlessly with regular traffic. In response to these challenges, machine learning is increasingly employed in NIDS. The advantage of machine learning-based approaches is their ability to learn and recognize complex patterns, often surpassing traditional signature-based methods. Such systems can adapt to evolving threats, reducing false positives and increasing detection rates. They are especially effective when previously unknown or zero-day threats emerge, as they can detect anomalies without relying on predefined signatures.

Recent work such as Chen et al. explored conversational features within the CTU-13 botnet dataset scenarios \cite{Chen}. They utilized multiple classifiers, such as Decision Tree, BayesNet, and the Random Forest Classifier. Velasco-Mata et al. delved into feature selection by employing the Information Gain and Gini Importance techniques \cite{velasco2021efficient}. Their efforts resulted in three pre-selected subsets containing five to seven features. On evaluating these subsets with Decision Tree, Random Forest, and k-Nearest Neighbors models, the Decision Trees with a five-feature set emerged as the top performers, achieving an impressive mean F1 score of 85\%. Dollah et al. focused their investigation on the detection of HTTP-based botnets \cite{Dollah}. They curated a labelled dataset by merging botnet traffic with genuine web browsing traffic. Their evaluation metrics spanned Accuracy, Precision, Recall, and FPR while utilizing classifiers like Decision Tree, k-nearest Neighbour, Naïve Bayes, and Random Forest. Remarkably, the k-Nearest Neighbours classifier stood out, registering an average accuracy of 92.93\%. Decision Tree's performance was superior to the Random Forest for the HTTP datasets they employed. Haddadi et al. Discriminatorally evaluated the performance of renowned IDS tools, Snort and BotHunter, against subsets from the CTU-13 dataset scenarios \cite{Haddadi}. Incorporating the C4.5 Decision Tree algorithm as their machine learning classifier, they experimented with feature sets derived from Tranalyzer and Argus network flow extraction tools. The Argus tool's features achieve an average DR of up to 99.45\%. Multilayer Transformer encoder and deep neural network is utilized in \cite{wu2024bot}. The Transformer encoder is utilized for encoding the implicit semantic relationship between the traffic bytes of the botnet and the deep learning network to capture the spatial relationship between the traffic bytes of the botnet, achieving 91.92\%  detection accuracy on the ISCX-2014 dataset. BotMark \cite{wang2020botmark} utilizes 15 statistical flow-based features and three graph-based features, applying k-means clustering to assess C-flows' similarity and leveraging least-square techniques and Local Outlier Factor (LOF) to evaluate graph-based anomalies. The model was tested with simulated network traffic from five recent botnets, including Mirai and Zeus, in a real computing environment, achieving an impressive 99.94\% detection accuracy. Shahhosseini et al. \cite{shahhosseini2022deep} proposes to extract only the header parts of packets, specifically from the transport (TCP/UDP) and network (IP) layers, avoiding the payload to ensure privacy and to handle the increasing use of encrypted traffic which makes the payload-based analysis less effective. The feature extraction is done using a Long Short Term Memory (LSTM) and uses a Random Forrest classifier to classify the botnets. Negative Sampling Algorithm Based on an Artificial Immune System is used in \cite{hosseini2022botnet} to reduce dimensionality by focusing on features that signify abnormal behaviour, effectively filtering out normal data, after which the data are scaled and using CNN and LSTM to classify the network traffic into Botnet or Normal.The Self-Attention mechanism is introduced in \cite{li2022botnet} where it uses Convolutional Block Attention Module (CBAM)-ResNet for spatial features and aids in understanding the sequence and temporal context. DCNNBiLSTM \cite{hnamte2023dcnnbilstm} utilizes a hybrid deep learning model combining Convolutional Neural Networks (CNN) and Bidirectional Long Short-Term Memory (BiLSTM) networks to detect botnet in Edge IoT devices. CNN layers first extract spatial features from the input data, which are then passed through BiLSTM layers that analyze temporal dependencies and sequences within the data. Similar work in HDLNIDS \cite{qazi2023hdlnids}  employs a deep-learning framework combining convolutional neural networks (CNNs) and recurrent neural networks (RNNs) where CNN is used to extract spatial features and RNN to analyze temporal features. Gunupdi et al. utilize \cite{kumar2024deep} Deep Residual Convolutional Neural Network (DCRNN) with Novel Binary Grasshopper Optimization Algorithm (NBGOA) as feature selector and Improved Gazelle Optimization Algorithm to search hyperparameter of the algorithm to reduce high false alarm rate of NIDS.

Machine learning and deep learning solutions have accurately detected malicious network flows. However, these algorithms have vulnerabilities, especially when faced with carefully crafted adversarial samples. This vulnerability is well-acknowledged in the field. Although there is a high output of scholarly activities from 2018 onwards on the topic of Intrusion Detection \cite{yaseen2023mapping}, a systematic review that explains the landscape of data poisoning attacks shows there is a lack of work that explicitly focuses on data poisoning attack \cite{aljanabi2023navigating}.
The challenge is even more pronounced in the context of Network Intrusion Detection Systems (NIDS). Unlike images, where adversarial perturbations might be straightforward, NIDS requires preserving syntactic and semantic relationships in the data. For instance, if one were to alter the 'duration' feature in a feature vector, a corresponding change in the 'rate' feature would be necessary to maintain coherence. Recent research has explored methods to generate adversarial features. Notably, the studies in \cite{alhajjar2021adversarial} and \cite{han2021evaluating} have employed Generative Adversarial Networks (GANs) to craft adversarial samples. Following the generation of these samples, the works in  \cite{alhajjar2021adversarial}, \cite{han2021evaluating}, and \cite{chen2020generating} have utilized Particle Swarm Optimization (PSO) to automate the search for optimal traffic mutants leveraging the capabilities of PSO. It has been observed in a recent study \cite{debicha2023adv} that even small perturbations in network traffic can bypass NIDS protections if strategically introduced. The study highlights that this can be achieved without compromising the core logic of the botnet attack. To ensure that the semantic relationship remains intact while modifying the feature vector, the approach taken in \cite{debicha2023adv} involves using a mask.
Additionally, they employ a sign-based adversarial attack, commonly called the 'sign attack'. Debicha et al. \cite{debicha2023tad}  leverages multiple strategically positioned adversarial detectors to improve detection rates over traditional single detector setups. Various transfer learning techniques, such as Feature Extraction and Fine-Tuning, as well as fusion rules like majority voting, Bayesian averaging, and Dempster-Shafer theory, are utilized, showcasing that a parallel IDS design with multiple detectors can better manage adversarial attacks. The practical feasibility of adversarial evasion attacks against machine learning-based NIDS is explored in Adv-Bot \cite{debicha2023adv}, demonstrating the potential to mislead these systems with strategically crafted adversarial botnet traffic. In their 2024 study, Roshan et al. \cite{roshan2024untargeted} assess the vulnerability of machine learning-based NIDS to white-box adversarial attacks and suggest heuristic defence strategies to increase system robustness. Employing advanced adversarial attack techniques like FGSM, JSMA, PGD, and C\&W, the study evaluates NIDS robustness. Meanwhile, Mohammadian et al. (2023) \cite{mohammadian2023gradient} focus on the susceptibility of deep learning-based NIDS to adversarial attacks, introducing a gradient-based method that identifies crucial features for creating adversarial perturbations using the Jacobian Saliency Map. This technique reduces the number of features needed, optimizing attack efficiency. Furthering this research, Roshan et al. (2023) \cite{roshan2023novel} deliver a detailed analysis of the efficacy of deep learning models in defending against adversarial attacks on network intrusion detection systems by employing robust adversarial attack algorithms and adversarial training as a defence strategy. Kumar et al. (2024) \cite{kumar2024generating} address vulnerabilities due to adversarial examples in DL/ML-based NIDS, introducing a variational autoencoder to create adversarial examples that not only evade detection but also comply with network security constraints. This approach achieves a 64.8\% success rate in evading DL/ML-based IDS. Lastly, a novel framework that integrates a Weighted Conditional Stepwise Adversarial Network (WCSAN) with Particle Swarm Optimization (PSO) is proposed by Barik et al. (2024) \cite{barik2024adversarial}. This framework enhances IDS by employing advanced feature selection techniques such as PCA and LASSO, improving detection accuracy through optimized adversarial training.

The framework across these studies introduces perturbations while preserving the network traffic's semantic and syntactic relationships. The extent of perturbation required to achieve desired outcomes has yet to be exhaustively analyzed in prior research. We aim to bridge this gap by delving deep into the nuances of perturbation levels in malware samples and evaluating their corresponding evasion capabilities. Furthermore, many adversarial samples are crafted using the gradient information from surrogate models. However, the transferability of these samples to non-gradient-based methods, such as Random Forests or Decision Trees, remains largely uncharted territory. We aim to shed light on this aspect by analyzing the transferability of generated adversarial samples to models without gradient information.
The need for reliable and trustworthy prediction systems in malware detection systems is becoming increasingly important. Recent work in the field of Android malware classification, such as the study by Barbero et al. \cite{barbero2022transcending}, has begun to explore the integration of conformal prediction to discard uncertain predictions, showcasing the potential of this methodology in enhancing prediction reliability. However, there is a notable gap in the current literature and application regarding integrating conformal prediction within the classification of network flow data. Our research aims to bridge this gap, shedding light on the potential of conformal prediction.

\section{Preliminaries}\label{Prelim-classifier}
This section describes the dataset, the feature extraction technique used, and the process of classification of network flow into Malware and Benign.

\subsection{Datasets}
In this research, we employed two distinct datasets: ISOT \cite{Saad2011DetectingPB} and ISCX\cite{BiglarBeigi2014EffectiveFS}. The ISOT dataset is a merged dataset of malicious and normal traffic datasets \cite{Saad}. The malicious traffic from the French chapter of the Honeynet Project comprises activity from the Storm, Waledac and Zeus botnets. The normal background traffic datasets are the product of merged datasets from Ericsson Research and the Lawrence Berkeley National Laboratory. This data includes traffic from HTTP web browsing, gaming and P2P clients. The ISCX Botnet dataset includes network traffic from a range of botnets such as Neris, Rbot, Virut, NSIS, SMTP, Zeus and normal activity traffic, which are captured by replaying over a network testbed topology\cite{Canadian}. The ISOT dataset has been divided into two parts, with a 70/30 split for training and testing purposes, respectively. From the training portion, we further reserve 5\% as a validation set to optimize the performance of the machine learning models. Similarly, for the ISCX dataset, although dedicated training and testing files were provided in the original publication, we also reserve 10\% of the training set as a validation set. We tune all our model's hyperparameters using the validation set. In our experiment, the random seed is set to 42 to ensure the experiments are reproducible.

\subsection{Feature Extraction and Feature Selection}\label{feature_selection}
In order to work with the ISOT and ISCX datasets, which are only available as .pcap files, feature extraction into network flow is necessary. Following the method used in \cite{Garcia}, we extracted 32 features and their corresponding labels. The extracted features are in Table \ref{table:features}. The complete descriptions of all the features for both the base and extended datasets can be found in the 'ra' man page \cite{Bullard}.

\begin{table}[!ht]
\centering
\footnotesize
\caption{Extracted features \cite{Bullard}}
\label{table:features}
\setlength{\tabcolsep}{3pt}
\begin{tabular}[c]{|p{60pt}|p{60pt}|p{60pt}|p{60pt}|}
\hline
Features & Features& Features & Features \\
\hline
SrcAddr &  dTtl& DstAddr &  TcpRtt\\
Proto  &  SynAck & Sport & AckDat\\
Dport & SrcPkts & State & DstPkts\\
sTos &  SAppBytes & dTos & DAppBytes\\
SrcWin & Dur & DstWin &  TotPkts\\
sHops  &  TotBytes & dHops &  Rate\\
LastTime & SrcRate & sTtl &  Label \\  

\hline
\end{tabular}
\end{table}

\begin{figure}[!ht]
    \centering
    \includegraphics[scale=0.45]{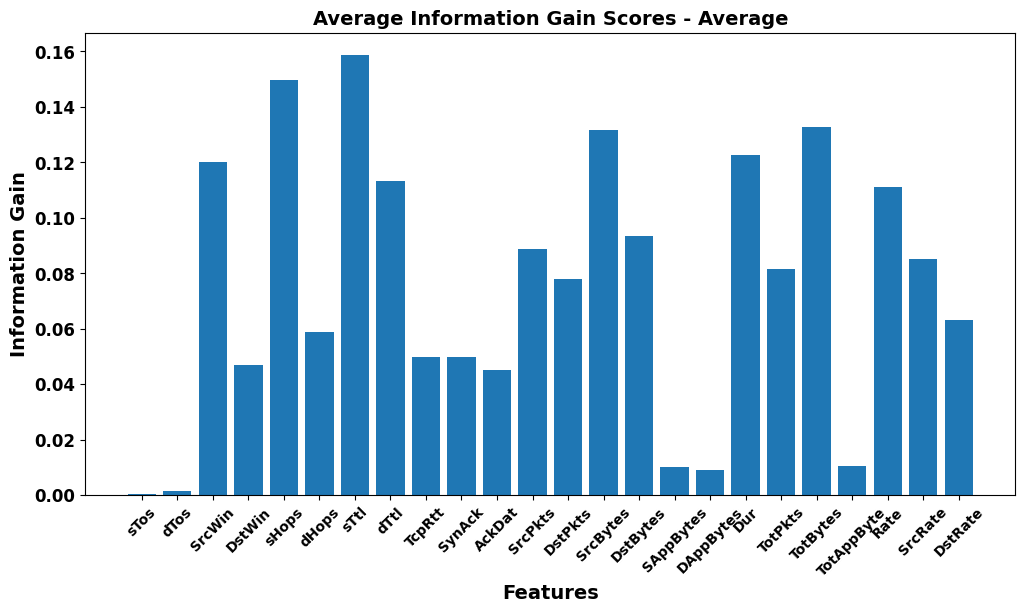}
    \caption{The average information gain of each feature, calculated from both the ISCX and ISOT datasets. }
    \label{fig:average_info_gain}
\end{figure}

\begin{figure}[!ht]
    \centering
    \includegraphics[scale=0.45]{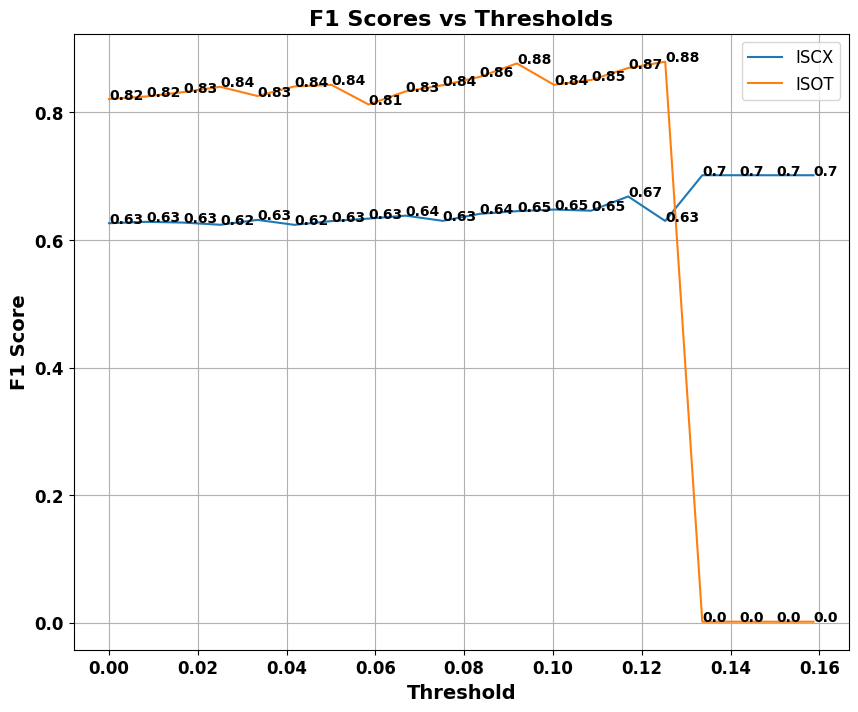}
    \caption{Comparative visualization of average F1 scores across various information gain thresholds for both ISCX and ISOT datasets}
    \label{fig:f1_thres_info_gain}
\end{figure}

We employed the Information Gain (IG) metric to determine the most relevant features. Information Gain (IG) is a statistical measure used in machine learning and information theory to determine how valuable a feature (or attribute) is in predicting the outcome or class of a given dataset. It measures the reduction in uncertainty or entropy achieved by partitioning a dataset based on that feature. In simpler terms, it quantifies the "gain" in our knowledge about the target variable after observing a feature.

Mathematically, the Information Gain for a feature 
A is defined as:

\begin{equation}
   IG(A) = H(D) - H(D|A)
\end{equation} 

Where $H(D)$ is the entropy of the dataset $D$  before the split. $H(D|A)$ is the expected entropy of the dataset $D$ after it has been partitioned based on the feature $A$. The average Information gain across the two datasets for each feature is shown in Figure \ref{fig:average_info_gain}. We then use a Random Forest classifier with default hyperparameters to get the F1 score against different Information Gain (IG) thresholds as shown in Figure \ref{fig:f1_thres_info_gain}. We aimed to identify the optimal IG threshold that maximizes the classification performance. From Figure \ref{fig:f1_thres_info_gain}, we select 0.09 as the threshold as it gives the best F1 score of 88 and 65 across the two datasets. The features selected using the threshold are SrcWin, sHops, sTtl, dTtl, SrcBytes, DstBytes, Dur, TotBytes and Rate. After feature extraction, we employ the Standard Scaler, which transforms the features to have a mean of 0 and a standard deviation of 1. This scaling process enhances the stability and speed of convergence during the model training phase, ensuring that no particular feature dominates due to its scale.

\subsection{Botnet Classification}
To detect botnet network flow and compare the results, we experimented with three classifiers: Decision Tree (DT), Random Forest (RF), and Neural Network (NN). Based on the feature selection described in \ref{feature_selection}, we used the following features to train the classifier: 'SrcWin', 'sHops', 'sTtl', 'dTtl', 'SrcBytes', 'DstBytes', 'Dur', 'TotBytes', and 'Rate'. We initially employed default hyperparameter configurations for DT and RF. We constructed a sequential model for the NN classifier consisting of six layers with nine units each, using the ReLU activation function. Following these were two layers with six units and ReLU activations, culminating in a final layer with a single unit employing the sigmoid activation function. The Adam optimizer and the binary cross-entropy loss function were used to train the neural network. We conducted training for fifty epochs with a batch size of 120. Table \ref{tab:default-hp-result} shows the initial results obtained using default hyperparameters. While these results provided valuable insights, we found that hyperparameter tuning was essential to enhance the model's predictive performance. Therefore, we employed three optimization algorithms, Random Search, Genetic Algorithm (GA) and Particle Swarm Optimization (PSO), to further optimize the hyperparameters of the DT, RF, and NN classifiers.

\begin{table}[!hbt]

\centering
\footnotesize
\caption{{\bf Classifiers performance with default hyper-parameters}}
\begin{tabular}{|l|l|l|l|l|l|l|l|l|l|l|l|l|l|l|}
\hline
& \multicolumn{4}{c|}{\bf ISCX} & \multicolumn{4}{c|}{\bf ISOT} \\ \hline
\textbf{Classifier} & \textbf{Acc} & \textbf{P} & \textbf{R} & \textbf{F1} & \textbf{Acc} & \textbf{P} & \textbf{R} & \textbf{F1}\\ \hline
DT & 73.7 & 74.59 & 69.29 & 71.84 & 99.24 & 98.21 & 91.7 & 94.89  \\ \hline
RF & 86.57 & 92.25 & 78.86 & 85.03 & 99.24 & 98.27 & 91.75 & 94.9  \\ \hline
NN & 64.87 & 60.52 & 78.66 & 68.4 & 98.75 & 97.71 & 91.84 & 94.68  \\ \hline
\end{tabular}
\begin{flushleft} Acc-Accuracy, P-Precision, R-Recall.
\end{flushleft}
\label{tab:default-hp-result}

\end{table}

\subsubsection{Hyper Parameter Optimization}
To optimize hyperparameters, we start with Random search, which operates by randomly selecting combinations of hyperparameters; we explore random search to establish a baseline performance level. We then explored nature-inspired algorithms, namely GA(Genetic Algorithm) and PSO (Particular Swarm Optimization). Nature-inspired algorithms are chosen for their robust exploration capabilities of the hyperparameter space through adaptive refinement. GA  adaptively focuses the search on regions of the hyperparameter space that have shown promise while still exploring new regions. Over generations, the population converges toward an optimal or near-optimal set of hyperparameters. PSO refines the search space by allowing particles to gravitate towards the best solutions found by themselves and by others in the swarm. This movement is influenced by individual and collective experiences, leading to convergence on optimal solutions as particles iteratively adjust their positions. The mechanism of exploration (searching new areas) and exploitation (improving known good areas) dynamically is crucial for effectively navigating the irregular and high-dimensional search spaces of hyperparameters, which makes them an ideal candidate for hyperparameter search.
\begin{itemize}

\item \textbf{Random Search(RS)}: Random Search optimization is a more efficient version of exhaustive search. In random search, a selection of parameter values is given and then iterates through them, picking randomly between the parameter values. After the optimization process, the best-performing collection of parameters is selected. Since stochastic values are chosen for each hyperparameter type from a set for a given classifier, this method is inefficient and may not result in an optimal hyperparameter set in a reasonable amount of computational time for guaranteed improvements \cite{Bergstra}. 

\item \textbf{Genetic Algorithm (GA)} is a search heuristic inspired by Charles Darwin's theory of natural evolution. Natural biological evolution tends to favour the fitter individuals in a population. They survive longer and breed, wherein the genes in the offspring are crossed over from the parents, allowing mutation in the genetic makeup. This natural selection process allows the desirable fittest characteristics to be propagated through the future population generations. We initialized a population of potential solutions, each representing a unique set of hyperparameters for the classifier. For example, in the Decision Tree, the hyperparameters such as criterion, splitter, max depth, and min samples split, among others, were chosen randomly from predefined ranges or sets. To evaluate the performance of each individual, we trained the machine learning classifier using the hyperparameters it represented and then assessed its performance on the validation set. We determined each individual's fitness using two key metrics: the F1 score and accuracy, with higher values indicating superior performance. To evolve our population, we employed a tournament selection mechanism, selecting individuals for crossover based on their fitness. We paired individuals to produce offspring using a uniform crossover method during the crossover phase. This method ensured that each hyperparameter from the parent individuals had an equal probability of being passed on to the offspring. We applied mutations after the crossover to introduce variability and maintain diversity in our population. In this mutation step, we randomly altered an individual's specific hyperparameters.
Over multiple generations, we repeated the selection, crossover, and mutation processes, allowing the population to evolve and improve its overall fitness. Upon the GA's completion, we used this best individual to train our classifier on the entire training dataset. We then evaluated the classifier's performance on the test data.

More formally, we initialized a population of size $P$, where each individual $ I_i $, for $i = 1, 2, ..., P $, is a vector of hyperparameters represented as $I_i = [h_1, h_2, ..., h_n] $, with $n $ being the total number of hyperparameters. To evaluate the performance of each individual, we employed a fitness function $( f(I_i) )$, which returns a tuple comprising the F1 score and accuracy, formulated as $f(I_i) = (F1(I_i), \text{Accuracy}(I_i))$. For the selection process, we used a tournament-based mechanism. A subset $S $ of size $T $ is randomly chosen from the population, and the individual $I^*$ with the highest fitness in $( S )$ is selected, denoted as $I^* = \text{argmax}_{I_i \in S} f(I_i) $. During the crossover phase, two-parent individuals, $ I_a $ and $( I_b )$, produce an offspring $I_c $. Each hyperparameter $( j )$ of the offspring is determined by $I_{c,j} = I_{a,j} $ with a probability of 0.5, or $I_{c,j} = I_{b,j} $ with a probability of 0.5. Mutation introduces variability. For each hyperparameter $h_j $ in an individual $I_i $, the hyperparameter might be replaced with a random value from its domain with a mutation probability $p_m $, or it remains unchanged. This iterative selection process, crossover, and mutation continue for $ G $ generations, where $ G $ is the predefined number of generations.

\item \textbf{Particle Swarm Optimization (PSO)} is a nature-inspired algorithm based on behaviour observed in swarms of fish, birds, etc. Each particle in the swarm represents a potential solution, encoded as a vector of hyperparameters, $ P_i = [h_1, h_2, ..., h_n] $, where $ n $ is the total number of hyperparameters. The position of each particle is updated iteratively based on its own best-known position and the global best-known position of the swarm. The equation governs the update rule for each particle's position:
$ P_{i, t+1} = P_{i, t} + v_{i, t+1} $
where $ v_{i, t+1} $ is the velocity of the particle at iteration $ t+1 $ and is computed as:
\begin{equation}
     v_{i, t+1} = w \cdot v_{i, t} + c_1 \cdot r_1 \cdot (p_{best} - P_{i, t}) + c_2 \cdot r_2 \cdot (g_{best} - P_{i, t})
\end{equation}
Here, $ w $ is the inertia weight, $ c_1 $ and $ c_2 $ are cognitive and social coefficients, respectively, $ r_1 $ and $ r_2 $ are random numbers between 0 and 1, $ p_{best} $ is the best-known position of the particle, and $ g_{best} $ is the best-known position among all particles in the swarm.

The fitness of each particle is evaluated using a combination of the F1 score and accuracy on the validation set. The PSO algorithm iteratively updates the positions of the particles to search for the optimal hyperparameters that maximize the fitness function. After a predefined number of iterations, the particle with the highest fitness is selected as the solution, representing our machine classifier's optimal set of hyperparameters. The classifier is then trained on the entire training set using these hyperparameters and evaluated on the test set to obtain the final performance metrics.

\end{itemize}

The hyperparameters of  Decision Tree, Random Forrest and Neural Network classifier were optimized using Random Search, Genetic Algorithm and Particle Swarm Optimization for comparison. The Decision Tree hyperparameter included the criterion, splitter, max depth, min samples per split, min samples per leaf, weight fraction leaf, max features, max-leaf nodes, impurity decrease, and the complexity parameter. We considered both "gini" and "entropy" for the criterion. We evaluated the splitter's "best" and "random" configurations. The max depth of the tree was varied from none to a maximum of 50, incrementing in steps of 3. We examined min samples per split from 2 up to 20 and min samples per leaf from 1 to 20. The weight fraction leaf was assessed from 0 up to 0.5, advancing in increments of 0.05. As for the max features, we explored "auto", "sqrt", "log2", and the possibility of it being none. The max-leaf nodes were checked in values ranging from none to 100, with a gap of 10 units between each. The impurity decrease hyperparameter was probed similarly to the weight fraction leaf, between 0 and 0.5 in 0.05 increments. Lastly, the complexity parameter, CCP alpha, was varied from 0 to 0.05, with 0.01 increments.

For Random Forrest, the hyperparameters include Number of estimators, Criterion, Maximum depth of the tree, Minimum number of samples required to split an internal node, Minimum number of samples required to be at a leaf node, Minimum weighted fraction of the sum total of weights required to be at a leaf node, number of features to consider when looking for the best split, Maximum number of leaf nodes, Threshold for early stopping in tree growth, Whether to use bootstrap samples and Complexity parameter. We set the number of estimators, or trees in the forest, by experimenting with values starting from 1 and going up to 191. For the criterion, the choices were "gini" and "entropy". The depth of individual trees, or max depth, was adjusted from 0 to a ceiling of 50, with intervals of 3. Samples required to split an internal node, or min samples split, were tested from 2 through 20. The minimal number of samples needed to be at a leaf node or min sample leaf, was explored from 1 to 20. The minimum weighted fraction of the total weights required to be at a leaf node was adjusted starting from 0 and reaching up to 0.5 in steps of 0.05.
Regarding the number of features to consider when looking for the best split or max features, we incorporated options like "sqrt", "log2", none, and even specific counts from 2 through 9. The maximum number of leaf nodes, or max-leaf nodes, was observed starting from no constraint and going up to 100, increasing in tens. The Threshold for early stopping in tree growth, or min impurity decrease, was set to range between 0 and 1, moving in increments of 0.1. The forest's bootstrap samples, determining whether to use out-of-bag samples to estimate the generalization score or not, were toggled between True and False. Lastly, the complexity parameter, CCP alpha, was allowed to vary from 0 up to 0.05, marking every 0.01 increment. For the neural network, we considered several hyperparameters. We selected the activation functions from 'relu', 'sigmoid', and 'tanh'. We chose the optimizers from 'adam', 'sgd', and 'rmsprop'. We picked the loss functions from 'binary\_crossentropy' and 'hinge'. We set the maximum number of hidden layers to 5 and allowed up to 50 nodes for each layer.

In the GA algorithm, the crossover mechanism was implemented using a uniform method, ensuring an even blend of genes from both parent entities. Regarding mutation, we randomly selected a hyperparameter and then altered its value. Selection was achieved through a tournament strategy, picking the finest out of five randomly chosen individuals based on their respective fitness scores. We employed a multi-objective approach to evaluate the fitness of each individual in the population. Specifically, we used the F1 score and accuracy of the classifier on the validation set as the fitness functions. The dual fitness measures ensured we did not just optimize for a single metric but considered the precision-recall balance (F1 score) and overall correct predictions (accuracy). We initiated the algorithm with a population of 100 individuals and ran it over 100 generations. After running the Genetic Algorithm, we identified the hyperparameters with the highest combined fitness score, and the best hyperparameter found is shown in Table \ref{tab:GA_hyper}. We then trained the Decision Tree classifier with these optimal hyperparameters on the entire training dataset and evaluated its performance on the test set. We reported key metrics such as accuracy, precision, recall, and F1 Score, which are shown in Table \ref{tab:ga-result}.

\begin{table}[!htb]
\centering
\footnotesize
\caption{{\bf Optimal hyperparameter configurations for various classifiers on ISCX and ISOT datasets, through Random Search optimization}}
\begin{tabular}{|l|l|p{12cm}|}
\hline
\textbf{Dataset} & \textbf{Classifier} & \textbf{Hyperparameters} \\ \hline
ISCX & DT & criterion: entropy, splitter: best, max\_depth: 24, min\_samples\_split: 16, min\_samples\_leaf: 20, min\_weight\_fraction\_leaf: 0, max\_features: log2, max\_leaf\_nodes: 100, min\_impurity\_decrease: 0.0, ccp\_alpha: 0.05 \\ \hline
& RF & n\_estimators: 1, criterion: entropy, max\_depth: 50, min\_samples\_split: 15, min\_samples\_leaf: 9, min\_weight\_fraction\_leaf: 0, max\_features: None, max\_leaf\_nodes: None, min\_impurity\_decrease: 0.0, bootstrap: False \\ \hline
& NN & num\_hidden: 4, nodes: 13, activation: tanh, optimizer: rmsprop, loss: binary\_crossentropy \\ \hline
\hline
ISOT & DT & criterion: gini, splitter: best, max\_depth: 45, min\_samples\_split: 15, min\_samples\_leaf: 12, min\_weight\_fraction\_leaf: 0.05, max\_features: auto, max\_leaf\_nodes: 40, min\_impurity\_decrease: 0.0, ccp\_alpha: 0.02 \\ \hline
& RF & n\_estimators: 1, criterion: entropy, max\_depth: 50, min\_samples\_split: 15, min\_samples\_leaf: 9, min\_weight\_fraction\_leaf: 0, max\_features: None, max\_leaf\_nodes: None, min\_impurity\_decrease: 0.0, bootstrap: False \\ \hline
& NN & num\_hidden: 5, nodes: 13, activation: tanh, optimizer: adam, loss: binary\_crossentropy \\ \hline
\end{tabular}
\label{tab:GA_hyper}
\end{table}

\begin{table}[!htb]
\centering
\footnotesize
\caption{{\bf Optimal hyperparameter configurations for various classifiers on ISCX and ISOT datasets, through GA optimization}}
\begin{tabular}{|l|l|p{12cm}|}
\hline
\textbf{Dataset} & \textbf{Classifier} & \textbf{Hyperparameters} \\ \hline
ISCX & DT & criterion: gini, splitter: random, max\_depth: 18, min\_samples\_split: 6, min\_samples\_leaf: 6, min\_weight\_fraction\_leaf: 0, max\_features: None, max\_leaf\_nodes: 90, min\_impurity\_decrease: 0.0, ccp\_alpha: 0.0 \\ \hline
& RF & n\_estimators: 2, criterion: gini, max\_depth: 27, min\_samples\_split: 2, min\_samples\_leaf: 13, min\_weight\_fraction\_leaf: 0.15, max\_features: log2, max\_leaf\_nodes: 60, min\_impurity\_decrease: 0.0, bootstrap: True, ccp\_alpha: 0.0 \\ \hline
& NN & num\_hidden: 5, nodes: 5, activation: tanh, optimizer: rmsprop, loss: hinge \\ \hline
\hline
ISOT & DT & criterion: entropy, splitter: best, max\_depth: 15, min\_samples\_split: 2, min\_samples\_leaf: 1, min\_weight\_fraction\_leaf: 0, max\_features: None, max\_leaf\_nodes: None, min\_impurity\_decrease: 0.0, ccp\_alpha: 0.0 \\ \hline
& RF & n\_estimators: 71, criterion: entropy, max\_depth: 42, min\_samples\_split: 3, min\_samples\_leaf: 2, min\_weight\_fraction\_leaf: 0, max\_features: 7, max\_leaf\_nodes: None, min\_impurity\_decrease: 0.0, bootstrap: True, ccp\_alpha: 0.0 \\ \hline
& NN & num\_hidden: 3, nodes: 29, activation: tanh, optimizer: rmsprop, loss: binary\_crossentropy \\ \hline
\end{tabular}
\label{tab:GA_hyper}

\end{table}

\begin{table}[!htb]

\centering
\footnotesize
\caption{{\bf Optimal hyperparameter configurations for various classifiers on ISCX and ISOT datasets, through PSO optimization}}
\begin{tabular}{|l|l|p{12cm}|}
\hline
\textbf{Dataset} & \textbf{Classifier} & \textbf{Hyperparameters} \\ \hline
ISCX & DT & criterion: entropy, splitter: random, max\_depth: 39, min\_samples\_split: 2, min\_samples\_leaf: 20, min\_weight\_fraction\_leaf: 0, max\_features: None, max\_leaf\_nodes: 20, min\_impurity\_decrease: 0.0, ccp\_alpha: 0.02 \\ \hline
& RF & criterion: entropy, splitter: random, max\_depth: 39, min\_samples\_split: 2, min\_samples\_leaf: 20, min\_weight\_fraction\_leaf: 0, max\_features: None, max\_leaf\_nodes: 20, min\_impurity\_decrease: 0.0, ccp\_alpha: 0.02 \\ \hline
& NN & activation: tanh, optimizer: sgd, loss: binary\_crossentropy, hidden\_layers: 1, layer\_nodes: 30, input\_nodes: 9 \\ \hline 
\hline
ISOT & DT & criterion: entropy, splitter: best, max\_depth: 30, min\_samples\_split: 16, min\_samples\_leaf: 11, min\_weight\_fraction\_leaf: 0, max\_features: log2, max\_leaf\_nodes: 80, min\_impurity\_decrease: 0.0, ccp\_alpha: 0.01 \\ \hline
& RF & n\_estimators: 61, criterion: entropy, max\_depth: 12, min\_samples\_split: 6, min\_samples\_leaf: 15, min\_weight\_fraction\_leaf: 0, max\_features: 4, max\_leaf\_nodes: 20, min\_impurity\_decrease: 0.0, bootstrap: True, ccp\_alpha: 0.0 \\ \hline
& NN & activation: relu, optimizer: adam, loss: binary\_crossentropy, hidden\_layers: 3, layer\_nodes: 40, input\_nodes: 9 \\ \hline
\end{tabular}
\label{tab:pso_hyperparams}

\end{table}

\begin{table}[!hbt]
\centering
\footnotesize
\caption{{\bf Classifiers performance with Random Search hyper-parameters}}
\begin{tabular}{|l|l|l|l|l|l|l|l|l|l|l|l|l|l|l|}
\hline
& \multicolumn{4}{c|}{\bf ISCX} & \multicolumn{4}{c|}{\bf ISOT} \\ \hline
\textbf{Classifier} & \textbf{Acc} & \textbf{P} & \textbf{R} & \textbf{F1} & \textbf{Acc} & \textbf{P} & \textbf{R} & \textbf{F1}\\ \hline
DT & 61.17 & 85.69 & 43.82 & 57.99 & 95.68 & 77.39 & 61.70 & 68.66  \\ \hline
RF & 71.35 & 80.87 & 69.64 & \textbf{74.83} & 99.23 & 98.12 & 91.76 & \textbf{94.84}  \\ \hline
NN & 76.39 & 90.73 & 68.39 & 77.99 & 98.48 & 94.35 & 85.34 & 89.62  \\ \hline
\end{tabular}
\begin{flushleft} 
Legend: Acc-Accuracy, P-Precision, R-Recall, F1-F1 score.
\end{flushleft}
\label{tab:rand-hp-result}
\end{table}

\begin{table}[!htb]
\centering
\footnotesize
\caption{{\bf Classifiers performance with GA optimized hyper-parameters}}
\begin{tabular}{|l|l|l|l|l|l|l|l|l|l|l|l|l|l|l|}
\hline
& \multicolumn{4}{c|}{\bf ISCX} & \multicolumn{4}{c|}{\bf ISOT} \\ \hline
\textbf{Classifier} & \textbf{Acc} & \textbf{P} & \textbf{R} & \textbf{F1} & \textbf{Acc} & \textbf{P} & \textbf{R} & \textbf{F1}\\ \hline
DT & 93.83 & 93.55 & 96.57 & \textbf{95.04} & 99.24 & 98.11 & 91.92 & 94.91  \\ \hline
RF & 84.02 & 89.57 & 83.61 & 86.49 & 99.27 & 98.16 & 92.3 & 
\textbf{95.14}\\ \hline
NN & 91.49 & 95.3 & 90.54 & 92.86 & 98.74 & 97.26 & 86.12 & 91.35\\ \hline
\end{tabular}
\begin{flushleft} Legend: Acc-Accuracy, P-Precision, R-Recall, F1-F1 score
\end{flushleft}
\label{tab:ga-result}

\end{table}

\begin{table}[!htb]
\centering
\footnotesize
\caption{{\bf Classifiers performance with PSO optimized hyper-parameters}}
\begin{tabular}{|l|l|l|l|l|l|l|l|l|l|l|l|l|l|l|}
\hline
& \multicolumn{4}{c|}{\bf ISCX} & \multicolumn{4}{c|}{\bf ISOT}  \\ \hline
\textbf{Classifier} & \textbf{Acc} & \textbf{P} & \textbf{R} & \textbf{F1} & \textbf{Acc} & \textbf{P} & \textbf{R} & \textbf{F1} \\ \hline
DT & 87.77 & 88.99 & 91.31 & \textbf{90.13} & 96.73 & 86.08 & 68.98 & 76.59 \\ \hline
RF & 80.62 & 80.71 & 89.77 & 85 & 98.74 & 99.26 & 84.21 & \textbf{91.12} \\ \hline
NN & 79.09 & 76.77 & 94.35 & 84.66 & 98.54 & 94.08 & 86.54 & 90.15 \\ \hline
\end{tabular}
\begin{flushleft} Legend: Acc-Accuracy, P-Precision, R-Recall, F1-F1 score
\end{flushleft}
\label{tab:pso-result}
\end{table}

The Particle Swarm Optimisation (PSO) used involved representing particles as vectors of indices that corresponded to classifier hyperparameters. The PSO dynamics were directed by cognitive (`c1`), social (`c2`), and inertia (`w`) coefficients, with values set at 1.5, 2, and 0.9, respectively. These coefficients helped to direct the particles towards individual and global best positions. Each particle's position corresponds to a specific combination of hyperparameters. To determine the fitness of each particle, we used a multi-objective approach, which involved evaluating the F1 score and accuracy of the classifier on the test set. Then, we transformed these fitness values into a single objective by computing the weighted sum of the F1 score and accuracy. Each objective contributed equally with a weightage of 50\%, ensuring that we optimised for a balanced trade-off between precision-recall (captured by the F1 score) and overall correct predictions (accuracy). We initialised the swarm with 100 particles and optimised over 50 iterations. Finally, we identified the global best particle position. We then used these optimal hyperparameters shown in Table \ref{tab:pso_hyperparams} to train the classifier on the entire training dataset. The performance of this classifier was subsequently assessed on the test set, and it is shown in Table \ref{tab:pso-result}. Finally, to establish a baseline and to show that the GA and PSO makes improvement on the perfromance we show the result of Random search in \ref{tab:rand-hp-result}.

\section{Adversarial Attack on Classification Model}\label{sec:Adv-attack}

In order to evaluate the robustness of our machine learning-based Network Intrusion Detection Systems (NIDSs), we crafted adversarial samples. Adversarial samples are instances of data that are intentionally perturbed in such a way as to deceive the model, leading to incorrect predictions. The best model in Section \ref{Prelim-classifier} acts as a classifier with no architectural information about the classifier to the attacker. We adopted the C\&W \cite{carlini2018audio} and GAN\cite{arjovsky2017wasserstein} attack, both renowned adversarial attack techniques, to evaluate the robustness of our model against adversarial perturbations on individual features. This section outlines the threat model, constraints associated with crafting these adversarial samples, the background of the C\&W and GAN attack and how we adapted it for our problem.

\subsection{Threat Model}
In our experimental setup, we operate under a grey-box attack scenario. Here, the attacker has complete knowledge of the dataset and features utilized by the model but lacks access to the model's parameters and architecture. This scenario is plausible, as in practical applications, features and datasets used in ML-based NIDSs are often disclosed through publications or documentation. Our primary objective is to scrutinize the vulnerabilities associated with specific features and investigate the transferability of adversarial samples across different models.
\subsection{Problem Definition}
Let $ N = \{N_{1}, N_{2}, ..., N_{m}\} $ be a dataset consisting of network traffic samples, where each sample $ N_i $ is characterized by a set of features $ \{F_1, F_2, ..., F_k\} $. The task at hand is to evaluate the robustness of a machine learning-based Network Intrusion Detection System (NIDS) model $ M: N \rightarrow \{0, 1\} $, which classifies traffic as benign ($ M(N_i) = 0 $) or malicious ($ M(N_i) = 1 $).

The focus is specifically on adversarial attacks targeting malicious network traffic samples with the intention of transforming them into adversarial samples that can evade detection by the NIDS model. For a given malicious sample $ N_i $ where $ M(N_i) = 1 $, the objective is to generate an adversarial counterpart $ N_i' $ that not only deceives the NIDS model into making an incorrect prediction, $ M(N_i') = 0 $, but also maintains plausible and coherent feature relationships to resemble legitimate network traffic, adhering to a set of constraints. To accomplish this, we utilize the Carlini \& Wagner (C\&W)\cite{carlini2018audio} method and GAN\cite{arjovsky2017wasserstein} method, adapting it to ensure that the perturbations applied to the malicious samples are subtle yet effective in evading detection while preserving the essential characteristics of legitimate network traffic. 

\subsection{C\&W Attack on the Classification model}
The primary goal of the C\&W attack is to find a perturbation that minimizes the distance between the original and perturbed input, 
while ensuring that the perturbed input is misclassified by the model. The objective function used in the C\&W attack is defined as:
\begin{equation}\label{eqn:c_w_objective}
   Objective = \|x' - x\|_2^2 + c \times f(x')
\end{equation}

$ x' $ is the adversarial example, $ x $ is the original input, and $ c $ is a constant that balances the trade-off between the perturbation magnitude and the misclassification. The function $ f(x') $ is designed to capture the misclassification condition and is defined as:
\begin{equation}
f(x') = \max \left( \max_{i \neq t} \left( Z(x')_i \right) - Z(x')_t, -k \right) 
\end{equation}
Here, $ Z(x') $ represents the logits or scores produced by the model for input $ x' $, $ t $ is the target class, and $ i $ iterates over all classes except the target class. The parameter $ k $ plays a pivotal role in determining the confidence with which the adversarial example should be misclassified.
The C\&W attack makes use of gradient-based optimization to craft the adversarial sample. Specifically, the gradient of the objective function to the input $ x' $ is computed. This gradient indicates how the value of the objective function will change for a small change in $ x' $. Mathematically:
   \begin{equation}
          \nabla_{x'} \text{Objective}
   \end{equation}

This gradient tells us the direction in which we should adjust $( x' )$ to decrease the value of the objective function most efficiently. The input $( x')$ is adjusted iteratively using the computed gradient. An optimizer is often used for this purpose. The following equation gives the update rule:
\begin{equation}
       x'_{new} = x'_{old} - \alpha \times \nabla_{x'} Objective
\end{equation}
Where $\alpha$ is the learning rate, which controls the step size during the optimization process, the gradient computation and the update rule are repeated until the Pre-defined maximum number of iterations is reached. The C\&W loss, which is the objective to be minimized during the attack, combines the L2 distance between the original and perturbed input with the misclassification condition. The L2 distance ensures that the adversarial perturbation is as tiny as possible, making it imperceptible to human observers. On the other hand, the term $ c \times f(x') $ ensures that the perturbed input $ x' $ is misclassified into the desired target class with a confidence determined by $ k $. The constant $ c $ is crucial as it determines the trade-off between these two objectives. The value of $k$ is essentially a margin or threshold that defines the confidence with which we want the adversarial example to be misclassified. If k were set to a positive value, the adversarial example would need to be misclassified and misclassified with a k margin of confidence. Positive $k$ value could lead to more significant perturbations, making the adversarial example further from the decision boundary and potentially more detectable. In our experiment, we set $k$ to 0, which means $ f(x') $ becomes positive (indicating a successful attack) as soon as the score for a wrong class exceeds the score for the correct class, even if just by a tiny margin. In our experiment, we have only benign and malware network flow, and we want malware network flow classified as benign; therefore, equation \ref{eqn:c_w_objective} can be written as shown in \ref{eqn:rewrite-objective} and \emph{Algorithm \ref{alg:objective_computation}} outlines the computation of this objective function.

\begin{equation}
\label{eqn:rewrite-objective}
 f(x') = max(Z(x')_{benign} - Z(x')_{malware},0)
\end{equation}

\begin{algorithm}[H]
\caption{Carlini-Wagner Adversarial Loss Calculation}
\label{alg:objective_computation}
\begin{algorithmic}[1]
\footnotesize
\State \textbf{Input:} $x, x', \text{model}, \text{target\_class}, c, \kappa$
\State \textbf{Output:} $\text{loss}$
\State \textbf{Description:} This function computes the Carlini-Wagner adversarial loss, which is a combination of the L2 distance between the original and perturbed inputs, and a confidence margin-based loss.

\Function{$f$}{$x', \text{model}, \text{target\_class}, \kappa$}
    \State $Z \leftarrow \text{model}(x')$ \Comment{Get model logits for perturbed input}
    \If{$\text{target\_class} = 0$}
        \State $Z_{\text{target}} \leftarrow 1 - Z[:, 0]$
        \State $Z_{\text{other}} \leftarrow Z[:, 0]$
    \Else
        \State $Z_{\text{target}} \leftarrow Z[:, 0]$
        \State $Z_{\text{other}} \leftarrow 1 - Z_{\text{target}}$
    \EndIf
    \State \textbf{return} $\max(Z_{\text{other}} - Z_{\text{target}}, -\kappa)$ \Comment{Compute confidence margin}
\EndFunction

\Function{cw\_loss}{$x, x', \text{model}, \text{target\_class}, c, \kappa$}
    \State $\text{l2\_dist} \leftarrow \text{reduce\_sum}((x - x')^2)$ \Comment{Calculate L2 distance}
    \State \textbf{return} $\text{l2\_dist} + c \cdot f(x', \text{model}, \text{target\_class}, \kappa)$ \Comment{Combine losses with balance parameter $c$}
\EndFunction

\State $\text{loss} \leftarrow \text{cw\_loss}(x, x', \text{model}, \text{target\_class}, c, \kappa)$ \Comment{Compute CW loss}
\State \textbf{return} $\text{loss}$
\end{algorithmic}
\end{algorithm}

In our experimental scenario, the classifier correctly identifies a malware network flow as malicious. As a result, the value of $ f(x') $ is zero, and its gradient for the objective function is negligible or absent. This condition presents a significant challenge in generating meaningful adversarial perturbations using gradient-based optimization techniques. When $ f(x') $ is zero, the optimization process becomes predominantly focused on minimizing the $ L_2 $ distance between the perturbed input $ x' $ and the original input $ x $. However, this approach might not yield an effective adversarial example, as it does not guarantee that the perturbed input will be misclassified.
To address the challenge of generating adversarial samples when the gradient of the objective function is negligible or zero, we introduce noise to the input, explicitly targeting the features we wish to attack. We generate random values between -1 and 1 and apply this noise element-wise to the attacked features. The noise is scaled relative to the original values of the attacked features, controlled by a specified magnitude. This scaling ensures that the perturbations are proportional to the original feature values, maintaining a balance that avoids excessive distortion. Algorithm \ref{alg:relative_noise} outlines the process of generating and applying this relative noise.

\begin{algorithm}[H]
\caption{Generate Relative Noise}
\label{alg:relative_noise}
\begin{algorithmic}[1]
\footnotesize
\State \textbf{Input:} x, feature\_mask, magnitude
\State \textbf{Output:} relative\_noise
\State \textbf{Description:} This function generates noise relative to the input $x$, scaled by the given magnitude, and masked by the feature\_mask, ensuring that only specific features are perturbed.
\State noise $\leftarrow$ random\_values\_in\_range(-1, 1, shape\_of(x)) \Comment{Generate random values}
\State relative\_noise $\leftarrow$ noise $\times$ x $\times$ magnitude \Comment{Scale noise relative to x}
\State relative\_noise $\leftarrow$ elementwise\_multiply(relative\_noise, feature\_mask) \Comment{Apply feature mask}
\State \textbf{return} relative\_noise
\end{algorithmic}
\end{algorithm}

The introduction of noise serves to nudge the input $ x' $ out of regions where the function $ f(x') $ is zero, facilitating a change in the model's classification. As a result, $ f(x') $ becomes positive, providing a gradient that can guide the optimization process effectively. With this gradient information, the optimization can minimize the $ L_2 $ distance between the original and perturbed input while ensuring that the input is misclassified. This approach is crucial for generating adversarial examples that are both effective (in terms of causing misclassification) and subtle (in terms of minimal perturbations). By targeting specific features for noise addition and scaling the noise relative to the original feature values, we ensure that the perturbations remain constrained and do not lead to overtly distorted adversarial examples.

\subsection{Generating Adversarial Network (GAN) attack:} The general architecture of GAN consists of a generator and a discriminator. In the context of fooling the NIDS, it also includes a substitute detector. The generator's (G) function creates synthetic data samples that mimic real network traffic. It starts from a random noise vector \( z \) and transforms it into a data sample \( x' \). Mathematically, the generator is represented by the following equation:

\begin{equation}
    G(z; \theta_g) = x'
\end{equation}
Here,  $\theta_g$ is the parameters of the generator. In our experiment, the generator is a neural network.

The Discriminator (D) role is to differentiate between real ( \( x \) ) and synthetic ( \( x' \) ) data samples. It outputs a probability score indicating the likelihood of a sample being real. The Discriminator is represented by the following equation:
\begin{equation}
    D(x; \theta_d) = p 
\end{equation}
Here, \( \theta_d \) are the parameters of the Discriminator, and \( p \) is the probability score. In our experiment, the Discriminator is a neural network. 

The Substitute Detector (Classifier, C) is a surrogate model for the actual NIDS model, classifying traffic as benign or malicious. It evaluates the synthetic samples \( x' \) and is represented as:
\begin{equation}
  C(x'; \theta_c) = y'   
\end{equation}
Here, \( \theta_c \) are the classifier's parameters, and \( y' \) is the predicted class.
The generator aims to produce samples that the Discriminator should classify as real, and the substitute detector should misclassify as malware. The loss function for the generator can be formulated as a combination of the Discriminator and the substitute detector's feedback. To train the generator, we use WGAN loss and a misclassification loss given by the substitute detector. It is formulated as 
\begin{equation}
\label{eqn:gen_loss}
  L_G = -\text{mean}(D(G(z))) + \text{mean}(C(G(z)))   
\end{equation}

The first term, \(-\text{mean}(D(G(z)))\), is the standard WGAN generator loss, encouraging the generator to create samples that the Discriminator evaluates as real. The second term, \(\text{mean}(C(G(z)))\), is the substitute loss, encouraging the generator to create samples that are classified as benign by the substitute detector. Here, minimizing the substitute loss means that the generator effectively tries to produce samples that the substitute detector scores as benign (closer to 0). The discriminator loss is designed to maximize the difference in evaluating real and synthetic samples. The loss is given by the following equation:

\begin{equation}
\label{eqn:discri_loss}
L_D = \text{mean}(D(G(z))) - \text{mean}(D(x))
\end{equation}

 The first Term $\text{mean}(D(G(z)))$ represents the average score assigned by the discriminatory to the generated (fake) samples. A higher score indicates that the Discriminator considers the fake sample to be more like real data. The second Term $\text{mean}(D(x))$ term represents the average score assigned by the Discriminator to the real data samples. \(x\) denotes the actual data. A higher score indicates that the Discriminator correctly recognizes the sample as real.    - Minimizing $\text{mean}(D(G(z)))$ pushes the Discriminator to assign lower scores to fake samples. Maximizing $\text{mean}(D(x))$ pushes the Discriminator to assign higher scores to real samples.

\subsection{Targeting Individual Feature}

Our attack focuses on attacking individual features without altering the entire data instance. A feature mask is employed to achieve these selective perturbations. A feature mask is a binary vector where each entry corresponds to a feature in the data. An entry with a value of 1 indicates that the corresponding feature can be modified, while 0 ensures that the feature remains untouched. This mask guides the adversarial crafting process, ensuring that only the desired features are susceptible to change. For instance, if the goal is to perturb the "Dur" (Duration) feature, the feature mask would have a one at the position corresponding to the "Dur" feature and 0 for all other positions. When the loss gradient to the input data is computed, this feature mask is applied, effectively zeroing out the gradients for all features except "Dur", which ensures that only the "Dur" feature is updated during the optimization process while all other features remain constant. The application of the feature mask is straightforward. After computing the gradients, they are element-wise multiplied with the feature mask. This operation ensures that gradients for untargeted features are nullified, preventing any changes during the optimization step.

\subsection{Constraint on Generating Adversarial sample}

After the optimization, it is essential to perturb a specific feature and ensure that related features are adjusted accordingly to maintain the coherence and plausibility of the traffic data. For instance, when the feature "Dur" (Duration) is perturbed, it is vital to adjust the "Rate" feature accordingly. The "Rate" is intrinsically linked to the "Duration" as it represents the number of bytes transferred per unit of time. If the "Duration" of a network session increases or decreases, the "Rate" at which data is transferred would naturally be affected. Specifically, if the "Duration" is shortened, the "Rate" would increase, indicating a faster data transfer, and vice versa. Similarly, when the "SrcBytes" (Source Bytes) feature is manipulated, it directly impacts the "TotBytes" (Total Bytes) feature. "TotBytes" is the sum of "SrcBytes" and "DstBytes" (Destination Bytes). Therefore, any change in "SrcBytes" necessitates a corresponding adjustment in "TotBytes". Additionally, to keep the "Rate" constant, the "Duration"  also need to be adjusted based on the new "SrcBytes" value. The features "sHops" and "dHops" represent the number of hop packets taken from the source to the destination. When these are altered, the Time-to-Live (TTL) values, represented by "sTtl" and "dTtl", must be adjusted. The TTL is typically initialized to a value (e.g., 255) and decremented at each hop. Thus, if "sHops" increases, "sTtl" would decrease, indicating that the packet has traversed more routers or switches. The relationship between the feature to be attacked and the necessary adjustment is shown in the following  Table \ref{tab:feature_relationships}.

\begin{table}[!ht]
\caption{Inter-feature relationships and adjustments post-perturbation}
\centering
\footnotesize
\begin{tabular}{|c|c|}
\hline
\textbf{Feature Perturbed} & \textbf{Related Adjustments} \\
\hline
Dur & Rate = $\frac{TotBytes}{Dur + \epsilon}$ \\
\hline
SrcBytes & $TotBytes = SrcBytes + DstBytes; Dur = \frac{TotBytes}{Rate}$ \\
\hline
DstBytes & $TotBytes = SrcBytes + DstBytes; Dur = \frac{TotBytes}{Rate}$ \\
\hline
TotBytes & 
\begin{tabular}{@{}c@{}} 
$SrcBytes = TotBytes - DstBytes;$ \\ 
$DstBytes = TotBytes - SrcBytes;$ \\
$Dur = \frac{TotBytes}{Rate + \epsilon}$ 
\end{tabular} \\
\hline
sHops & $sTtl = \text{INITIAL TTL} - sHops$ \\
\hline
sTtl or dTtl & $sHops = INITIAL TTL - sTtl$ \\
\hline
Rate & $Dur = \frac{TotBytes}{Rate + \epsilon}$ \\
\hline
SrcWin & No adjustment needed since it is not dependent on other features \\
\hline
\end{tabular}

\label{tab:feature_relationships}
\end{table}

\subsection{Defense Against Adversarial Sample}
To protect against the Adversarial sample generated, we use Adversarial retraining. Adversarial retraining is a defence strategy adopted to enhance the robustness of machine learning models against adversarial attacks. At its core, adversarial retraining involves incorporating adversarial samples into the training dataset and retraining the model. By doing this, the model learns from genuine data and adversarial perturbations. The hope is that this dual exposure during training will equip the model to better recognize and thwart adversarial attempts during actual deployment.

\section{Conformal Prediction in Network Flow Detection} \label{conformal_prediction}

Reliable and confident prediction is an essential aspect of network intrusion detection systems (NIDS), where the stakes are high and the margin for error is minimal. In this context, the conformal prediction framework, as introduced by Vovk et al. \cite{vovk2005algorithmic}, offers a robust and principled approach to not only classify network traffic but also quantify the certainty of each prediction. This section delves into the conformal prediction method, which has been adapted for application in network flow detection. For a comprehensive explanation, refer to Vovk et al.'s original paper \cite{vovk2005algorithmic}.

When analyzing a dataset of network flows, we divide it into training, calibration, and test sets. Using the training set, we train a classifier, represented as $ \hat{f} $. For each input instance $ x $ (symbolizing the features of a network flow), the classifier $\hat{f}$ yields two probabilities: $ \hat{f}_0(x) $ and $\hat{f}_1(x)$. The former indicates the likelihood that the instance $x$ represents normal traffic, while the latter signifies the probability of it being malicious. These probabilities, confined within the [0,1] interval, always sum to 1. We employ the calibration set to derive conformal scores $s_i=1-\hat{f}\left(X_i\right)_{Y_i}$, where $s_i$ stands for one minus the predicted probabilities of the original class, with $Y_i$ representing the true label of the instance (0 for normal and 1 for malicious). When $\hat{f}$ correctly predicts the class, $s_i$ remains low. In contrast, incorrect predictions result in a high $s_i$ value. Subsequently, we determine a threshold $\hat{q}$ from the $s_i$ values in the calibration set. This threshold represents the $\lceil(n+1)(1-\alpha)\rceil/n$ empirical quantile of the scores from $s_1$ through $s_n$. For fresh instances in the test set, we formulate a prediction set $C(X_{test})=\left\{y:\hat{f}(X_{test})_y\ge 1-\hat{q}\right\}$. This prediction set satisfies the following condition, given adequate trials \cite{vovk2005algorithmic}:

\begin{equation}
\label{coverage}
    1-\alpha\le P(Y_{test}\in C(X_{test}))
\end{equation}

Here, $X_{test}$ and $Y_{test}$ denote a test data point from a similar distribution, while the user specifies $\alpha$ as the desired error rate within [0,1]. The likelihood of the prediction set containing the correct label is at least $1-\alpha$—a measure known as \textit{coverage}. For instance, with $\alpha$ set at 0.5, the prediction set has a minimum 95\% 
Upon evaluating a network flow instance from the dataset, the function $C$ outputs a set based on the classifier's probability estimates. In the binary classification context of network traffic as normal or malicious, $C$ can yield one of the following outputs: (a) $\left\{ \emptyset \right\}$ (b) $\left\{ Normal \right\}$ (c) $\left\{ Malicious \right\}$ or (d) $\left\{ Normal, Malicious \right\}$. For instance, considering $\hat{q}$ as 0.08, and given an input $x_i$ that produces probabilities of 0.95 for malicious and 0.05 for normal traffic, the function $C$ will return $\left\{ Malicious \right\}$. If the probabilities were 0.4 for malicious and 0.6 for normal, with $\hat{q}$ still at 0.08, then $C$ would yield $\left\{ \emptyset \right\}$. In binary classifications, the eventuality of $C$ producing an output of $\left\{ Normal, Malicious \right\}$ is improbable due to the sum of probabilities being 1 and a $\hat{q}$ value of 0.5. A $\hat{q}$ of 0.5 indicates a poorly trained classifier, limiting the possible outputs of $C$ to $\left\{ \emptyset \right\}$, $\left\{ Normal \right\}$, or $\left\{ Malicious \right\}$. If a network flow instance receives an output of $\left\{ \emptyset \right\}$, we should reject the classifier's classification, potentially marking it for further review. Conversely, outputs of $\left\{ Normal \right\}$ or $\left\{ Malicious \right\}$ indicate accepted classifications, backed by the coverage assurance from equation \ref{coverage}. For an in-depth tutorial, please refer \cite{angelopoulos2021gentle}.

\section{Evaluation}

\subsection{Experimental Setup and Results for Crafting Adversarial Sample}

To ensure a comprehensive evaluation of the robustness of our models against adversarial attacks, we meticulously crafted our experimental procedure.
Our approach utilizes a surrogate model, a neural network we have pre-trained and discussed in the preliminaries section. This surrogate model serves as a substitute for the target model during the adversarial crafting process, providing gradient information that guides the generation of adversarial samples. Initially, we identify instances correctly predicted as malware by our surrogate model for each dataset ISCX and ISOT. By filtering the dataset based on these criteria, we concentrated our adversarial efforts on the samples most confidently identified by our model, thereby offering a test for our model's resilience. The attack was performed using the C\&W and GAN adversarial attack method described in Section \ref{sec:Adv-attack}. We employed Algorithm \ref{alg:generate_cw} to iteratively generate adversarial examples using C\&W attack and Algorithm \ref{alg:gan_training_feature_mask} for GAN attack. This technique was iteratively applied to each of our selected features, namely 'SrcWin', 'sHops', 'sTtl', 'dTtl', 'SrcBytes', 'DstBytes', 'Dur', 'TotBytes', and 'Rate'. This algorithm ensures that only specific features are perturbed, maintaining the semantic integrity of the data. The process involves initializing with relative noise in C\&W and generation of sample from random noise in GAN, applying a feature mask, and then optimizing the objective function to craft the adversarial examples. The primary objective was to subtly manipulate these samples so our model would misclassify them benign using the objective function shown in equation \ref{eqn:c_w_objective} and \ref{eqn:gen_loss}.

\begin{algorithm}
\caption{Generate CW Adversary}
\label{alg:generate_cw}
\begin{algorithmic}[1]
\footnotesize
\State \textbf{Input:} model, x, target\_class, feature\_mask, c, epsilon, iterations, clip\_min, clip\_max
\State \textbf{Output:} x\_prime
\State \textbf{Description:} This function generates an adversarial example by iteratively applying perturbations to the input $x$. The perturbations are guided by the CW loss and are masked to affect only certain features.
\State noise $\leftarrow$ GenerateRelativeNoise(x, feature\_mask) \Comment{Initialize with relative noise}
\State x\_prime $\leftarrow$ x + noise \Comment{Apply initial perturbation}
\State learning\_rate $\leftarrow$ epsilon \Comment{Set learning rate}
\For{iteration $\leftarrow$ 1 to iterations}
    \State loss $\leftarrow$ CalculateCwLoss(x, x\_prime, model, target\_class, c) \Comment{Calculate CW loss}
    \State gradient $\leftarrow$ compute\_gradient(loss, x\_prime) \Comment{Compute gradient of loss w.r.t x\_prime}
    \State masked\_gradient $\leftarrow$ elementwise\_multiply(gradient, feature\_mask) \Comment{Apply feature mask to gradient}
    \State x\_prime $\leftarrow$ x\_prime - learning\_rate $\times$ masked\_gradient \Comment{Update x\_prime}
    \State x\_prime[feature\_mask] $\leftarrow$ clip\_values(x\_prime[feature\_mask], clip\_min, clip\_max) \Comment{Clip values to stay within bounds}
\EndFor
\State \textbf{return} x\_prime
\end{algorithmic}
\end{algorithm}

\begin{algorithm}[H]
\caption{Training Generative Adversarial Network for NIDS Evasion}
\label{alg:gan_training_feature_mask}
\begin{algorithmic}[1]
\footnotesize
\State \textbf{Input:} real\_data, batch\_size, latent\_dim, Discriminator\_iterations, generator\_iterations, feature\_mask, scaler, feature\_min, feature\_max
\State \textbf{Output:} trained\_generator, trained\_Discriminator

\State \textbf{Initialize:} generator, Discriminator

\For{each epoch}
    \For{each batch in real\_data}
        \State $X_{real} \gets$ Sample real data
        \State $z \gets$ Sample random noise of shape (batch\_size, latent\_dim)
        \State $X_{fake\_raw} \gets$ generator(z) \Comment{Generate fake data}
        \State $X_{fake} \gets$ Apply feature mask to $X_{fake\_raw}$, clip to [feature\_min, feature\_max]

        \For{c\_iter from 1 to Discriminator\_iterations}
            \State $fake\_pred \gets$ Discriminator($X_{fake}$)
            \State $real\_pred \gets$ Discriminator($X_{real}$)
            \State $c\_loss \gets$ Discriminator\_loss($real\_pred$, $fake\_pred$)
            \State Update Discriminator weights to minimize $c\_loss$
        \EndFor
        
        \For{g\_iter from 1 to generator\_iterations}
            \State $z \gets$ Sample random noise of shape (batch\_size, latent\_dim)
            \State $X_{fake\_raw} \gets$ generator(z) \Comment{Generate new fake data}
            \State $X_{fake} \gets$ Apply feature mask to $X_{fake\_raw}$, adjust with scaler, clip to [feature\_min, feature\_max]
            \State $fake\_pred \gets$ Discriminator($X_{fake}$)
            \State $substitute\_pred \gets$ substitute\_detector($X_{fake}$)
            \State $g\_loss \gets$ generator\_loss($fake\_pred$, $substitute\_pred$)
            \State Update generator weights to minimize $g\_loss$
        \EndFor
    \EndFor
\EndFor

\State \textbf{return} generator, Discriminator
\end{algorithmic}
\end{algorithm}

For C\&W attack, given the size of malware samples identified by our surrogate model initially, generating adversarial samples for the entire malware samples at once could be computationally intensive. Hence, we adopted a batching approach for the generation. We defined batches with a BATCH\_SIZE of 10,000, dividing our dataset into manageable chunks. To commence the attack, we selected a feature and added some noise. To generate the noise, we generate random values within the range of -1 to 1 for each element in the batch. The rationale behind this randomness is to provide an unpredictable but constrained starting point for adversarial perturbations. Instead of relying solely on the random values, the noise was made 'relative' by multiplying it with the original data values, the relative value 0.1(10\%) and the noise to ensure that features with larger magnitudes have proportionally larger noise values, making the noise relative to the magnitude of the feature it perturbs and also that the adversarial examples start close in the feature space to the genuine data points. The final step in the noise generation was the application of a feature mask. This mask, essentially a binary vector, was employed to selectively activate the perturbation on specific features of interest while keeping other features untouched. By multiplying the relative noise with this feature mask, we effectively zeroed out the noise for features we intended to preserve. This ensured that the adversarial crafting only impacted the features we wanted to target. We then optimize iteratively the objective in equation \ref{eqn:c_w_objective} to generate adversarial examples using Adam optimizer with a learning rate set to 0.0001 and the constant  $c$ of the objective function, which determines the trade-off between the perturbation's magnitude and the classification error set to 0.01. In each iteration, we computed the gradient of this objective function concerning the perturbed input. However, this gradient is masked so that updates only occur in the desired feature direction, ensuring the semantic and syntactic relationships between features are maintained.
Post gradient computation, we updated our adversarial example in the direction that would increase the classification error. It is worth noting that after every iteration, we imposed constraints on the feature modified in our adversarial example to the min and max of the feature in the dataset to ensure its values did not breach the min and max of that particular feature in the dataset. This was crucial for retaining the realism and legitimacy of the crafted examples. The process mentioned above was iteratively repeated until either the adversarial example was crafted satisfactorily or a set threshold of iterations was reached, which is set to 2000 in our experiment.

\begin{algorithm}[H]
\caption{CW Batch}
\label{alg:C_w_batch}
\begin{algorithmic}[1]
\footnotesize
\State \textbf{Input:} model, scaler, input\_samples, target\_class, 
\State \hspace{1.1cm} feature\_name, feature\_min, feature\_max, 
\State \hspace{1.1cm} it\_value, c
\State \textbf{Output:} perturbed\_samples
\State \textbf{Description:} This algorithm generates a batch of adversarial examples using the Carlini-Wagner method. It adjusts features according to a given feature mask, and ensures that the dependencies between features are maintained.
\State target\_labels $\gets$ zeros\_like((input\_samples)) \Comment{Initialize target labels as zeros}
\State original $\gets$ inverse\_transform(scaler, input\_samples) \Comment{Transform inputs back to original scale(since the input is in standard sclae)}
\State feature\_index $\gets$ index(feature\_name in feature\_list) \Comment{Get index of the feature to be perturbed}
\State feature\_mask $\gets$ [1 if i == feature\_index else 0 for i in range(9)] \Comment{Create mask for selected feature}
\State perturbed\_samples $\gets$ GenerateCwAdversary(model, input\_samples, 
\State \hspace{2.5cm} target\_class, feature\_mask, c, 0.001, it\_value, 
\State \hspace{2.5cm} feature\_min, feature\_max) \Comment{Generate adversarial examples}
\State perturbed\_samples\_original $\gets$ inverse\_transform(scaler, perturbed\_samples) \Comment{Transform perturbed samples back to original scale}
\State perturbed\_samples\_original[:, feature\_index] $\gets$ clip\_values(
\State \hspace{2.5cm} perturbed\_samples\_original[:, feature\_index], feature\_min, feature\_max) \Comment{Clip the feature values to be within valid range of the train dataset}
\State AdjustDependencies(perturbed\_samples\_original, feature\_name, feature\_index) \Comment{Adjust dependencies between features}
\State perturbed\_samples $\gets$ transform(scaler, perturbed\_samples\_original) \Comment{Scale perturbed samples back}
\State \textbf{return} perturbed\_samples \Comment{Return the generated adversarial examples}
\end{algorithmic}
\end{algorithm}
As mentioned before, generating these adversarial examples was not just a random tweaking of feature values. Instead, we maintained the semantic and syntactic relationship between the features. We began by isolating the specific feature we intended to perturb. Once identified, we then applied the C\&W method. After generating the adversarial samples, the subsequent step ensured that these perturbations maintained the inter-feature relationships. For instance, if the 'Dur' feature was modified, adjustments were made to the 'Rate' feature to preserve their natural relationship. Likewise, changes to the 'SrcBytes' would cascade to both the 'TotBytes' and 'Dur' features to uphold the integrity of the data. Similarly, any perturbation to features like 'sHops' would subsequently impact 'sTtl' and vice versa. The relationship is detailed in Table \ref{tab:feature_relationships}. Algorithm \ref{alg:C_w_batch} outlines the entire process of generating adversarial samples in batches 



%

\begin{figure*}
    \centering
    \includegraphics[scale=0.16]{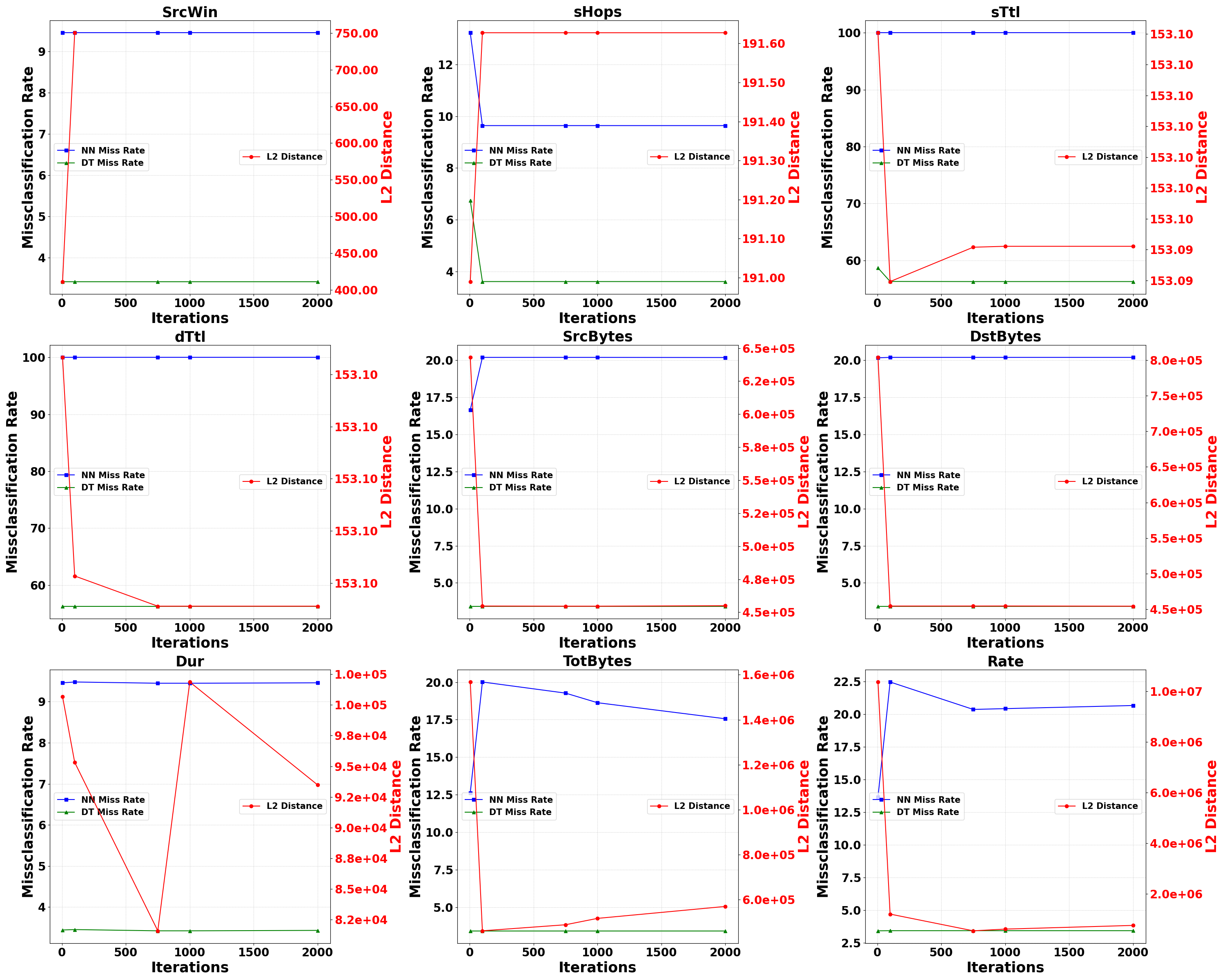}
    \caption{Average L2 distance and misclassification rate in ISCX dataset using C\&W attack}
    \label{fig:iscx-l2_C&W}

    \centering
    \includegraphics[scale=0.16]{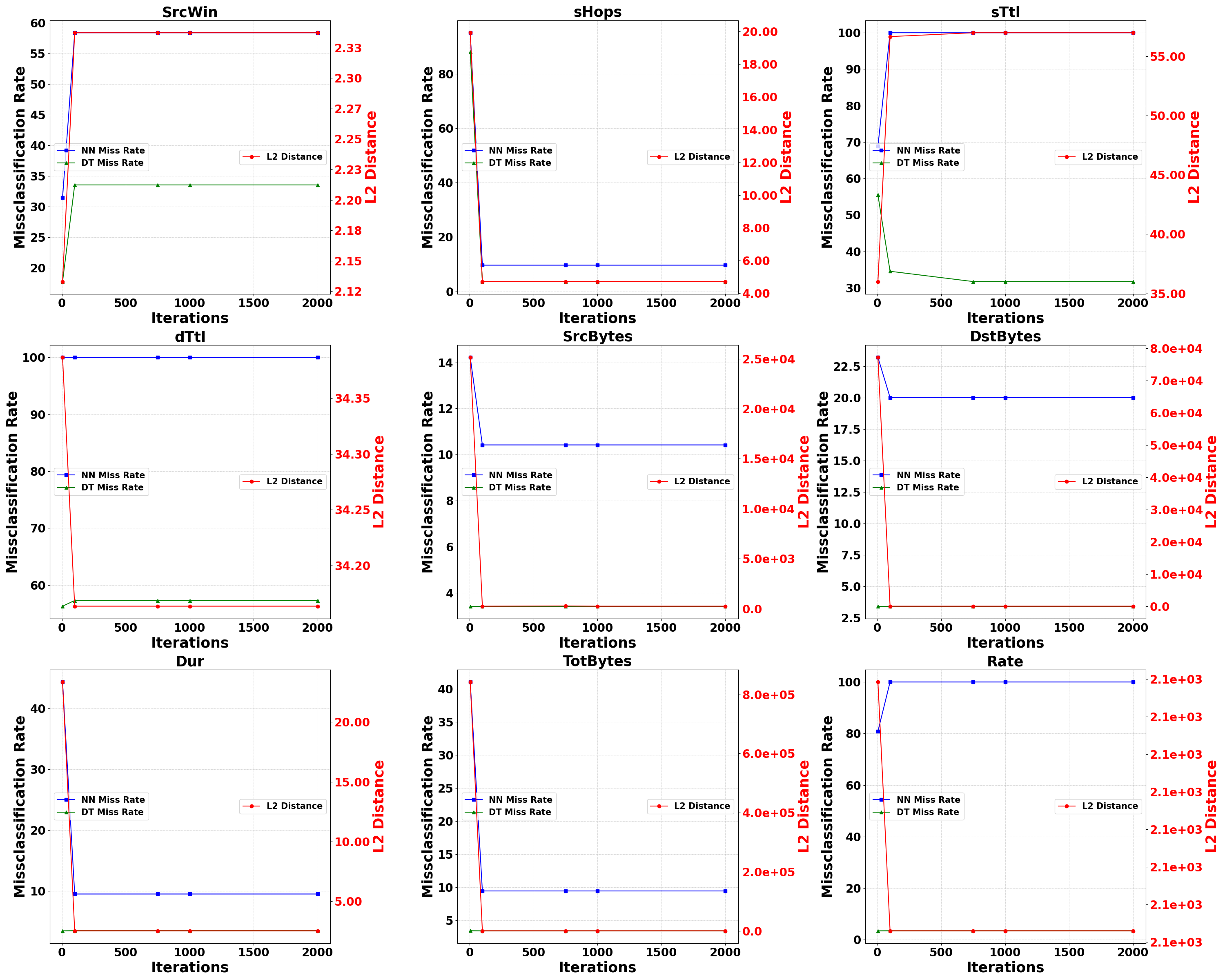}
    \caption{Average L2 distance and misclassification rate in ISCX dataset using GAN attack}
    \label{fig:iscx-l2_gan}
\end{figure*}

\begin{figure*}
    \centering
    \includegraphics[scale=0.16]{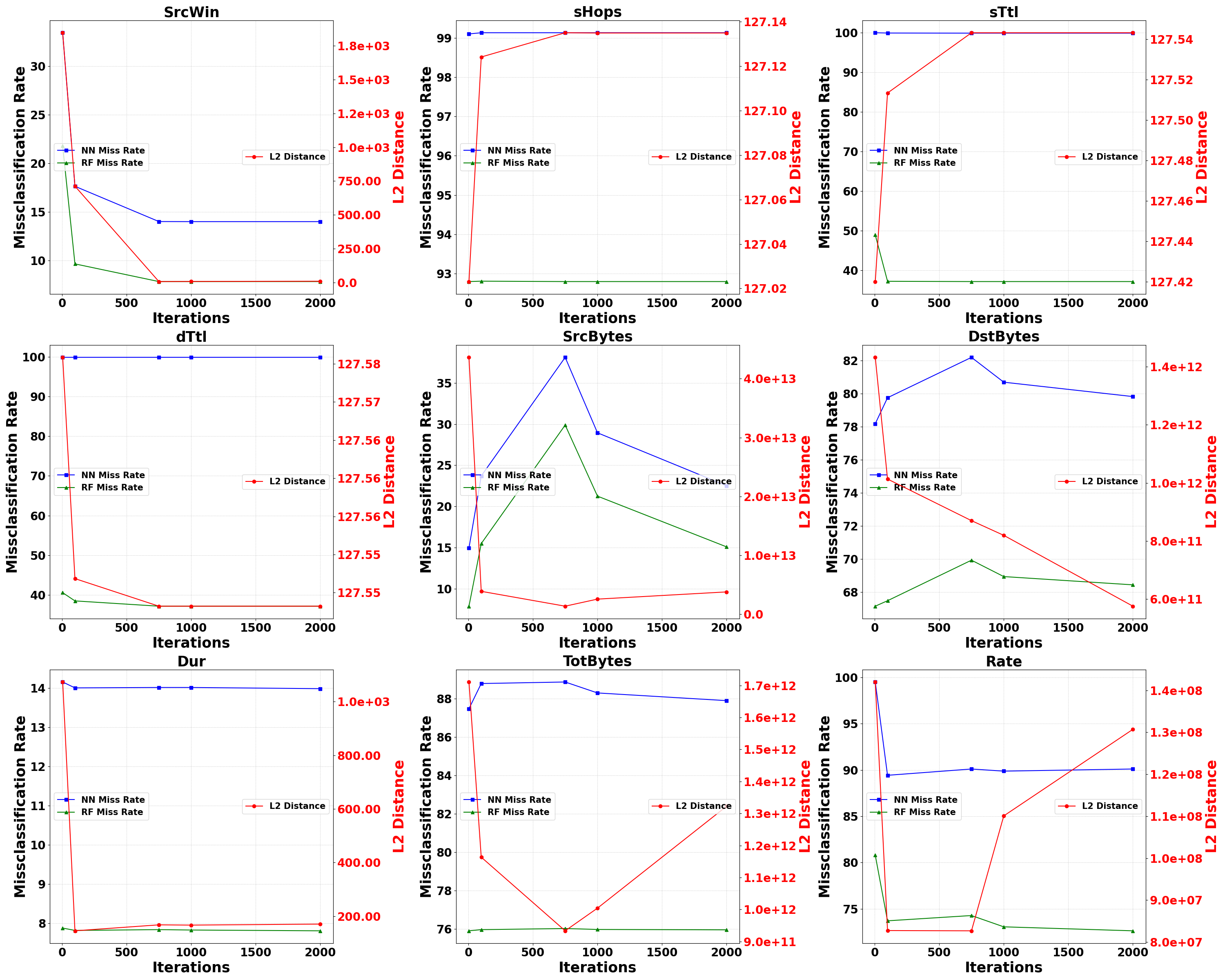}
    \caption{Average L2 distance and misclassification rate in ISOT dataset using C\&W attack}
    \label{fig:isot-l2_C&W}

    \centering
    \includegraphics[scale=0.16]{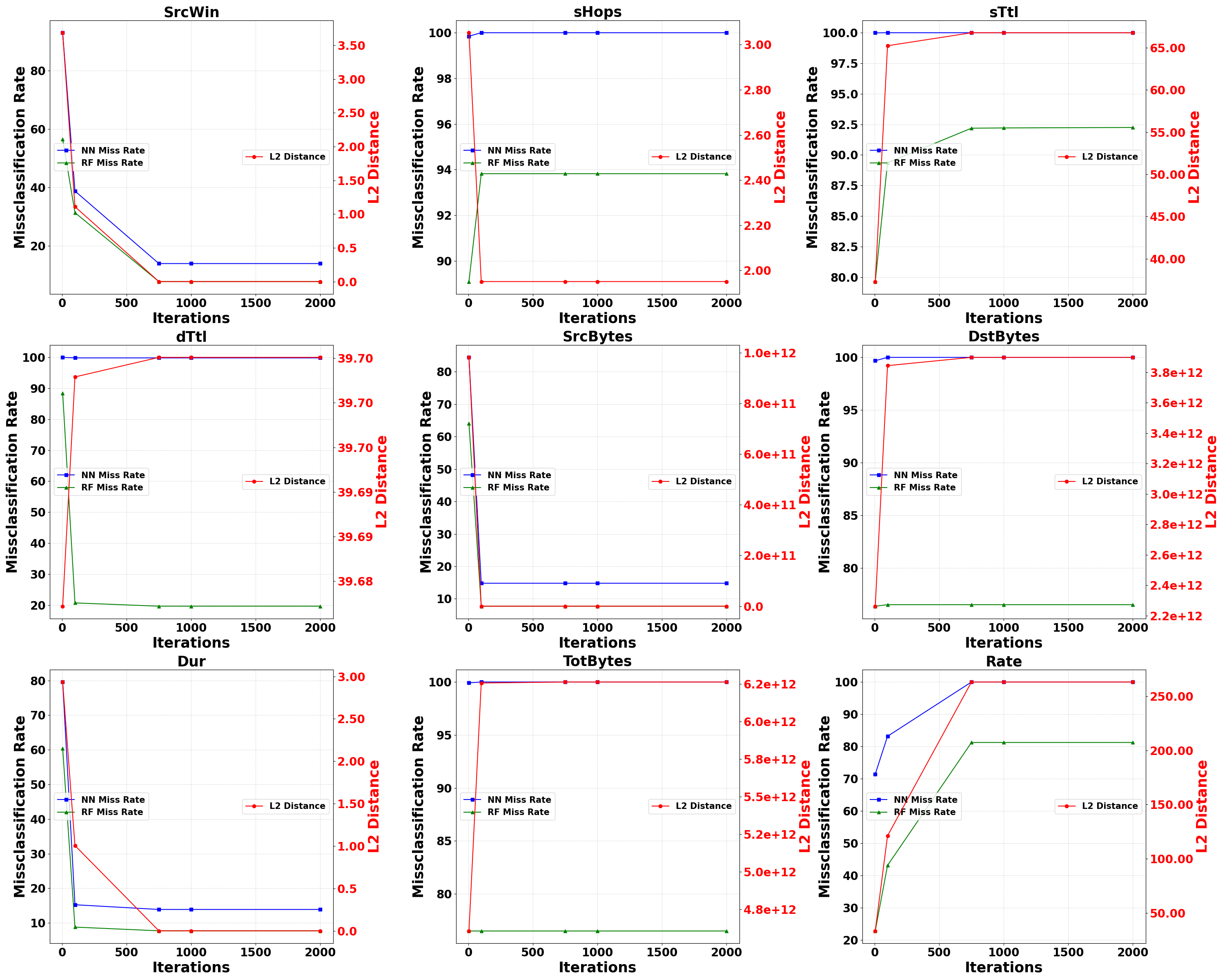}
    \caption{Average L2 distance and misclassification rate in ISOT dataset using GAN attack}
    \label{fig:isot-l2_gan}
\end{figure*}

In GAN attack, the process of adversarial sample uses a generator-discriminator architecture along with a surrogate model. The generator's role was to create synthetic data samples while the discriminator evaluated these samples' authenticity. The substitute detector, a neural network, served to simulate the target model's response to the adversarial inputs. Unlike the C\&W, the adversarial generation begins with the creation of random latent vectors drawn from a normal distribution. The dimension of the latent variable is set to 100, from which the generator learns to map to the data space.

The generator is a neural network with two layers. The input layer is a dense layer with 128 units. It uses the ReLU  (Rectified Linear Unit) activation function. The output layer is a dense layer with ten units, which is the number of features that we selected and uses the tanh activation function. To ensure that the output of the generator matches the scale of the original data, especially after standard normalization processes like StandardScaler, we apply a scaling transformation. This transformation involves multiplying the output by a scaling factor, which is the range of the data divided by 2, and then adding an offset, which is the average of the data's minimum and maximum values. This step is crucial for maintaining the realism of the generated data. The discriminator is another neural network that evaluates the authenticity of both real and synthetic data samples. The discriminator input is a dense layer with 128 units and a ReLU activation function. The final layer of the discriminator is a single-unit dense layer, which outputs a score representing the discriminator's assessment of how real or fake the input data appears. In standard GANs, the discriminator's output logit is often passed through a sigmoid activation function during the training phase to calculate the binary cross-entropy loss. However, the raw logit is used directly in WGANs and certain other GAN variants. This approach is based on the principle that using the raw score can lead to more stable training dynamics and better convergence properties, as it avoids potential issues related to vanishing or exploding gradients that can occur with bounded activation functions. The discriminator's loss is shown in equation \ref{eqn:discri_loss}.

The surrogate model, which is a neural network train in our preliminary section, acts as a proxy for the actual target model that the adversarial examples are intended to deceive. In the adversarial crafting process, the generator is tasked with producing samples that can fool both the discriminator and the surrogate model. The generator's training involves optimizing its ability to produce samples that are both realistic (as judged by the discriminator) and deceptive (as judged by the surrogate model). The loss is given in equation \ref{eqn:gen_loss}, which is designed to achieve a balance between producing samples that appear authentic to the discriminator and deceiving the surrogate model. During the optimization process, we employed the Adam optimizer with a learning rate of 0.0001. Our approach to adversarial example generation began with the selection of a target feature from the set 'TotAppByte', 'SAppBytes', 'DAppBytes', 'DstBytes', 'TotBytes', 'SrcBytes'. Similar to C\&W, we use a binary vector to zero out the gradient of other features and only change the target feature and finally adjust related features using the relationship detailed in Table \ref{tab:feature_relationships}.

To further our understanding regarding the change in the original value compared to the adversarial sample in different points of iteration, we considered multiple iteration checkpoints, which are 5,100,750,1000,2000. For every targeted feature and iteration value, we collected the adversarial samples. Considering the nuances of adversarial attacks, it is not just about how often the model is deceived but also about the magnitude or intensity of the changes made to the original data. For this reason, we determined the average L2 distance difference between the original and adversarial data points, offering a quantitative measure of these adversarial perturbations. This relationship, detailing the trade-off between the difference in L2 distance of adversarial modifications with the original sample and their effectiveness, is captured in Figures  \ref{fig:iscx-l2_C&W}, \ref{fig:iscx-l2_gan},\ref{fig:isot-l2_C&W} and \ref{fig:isot-l2_gan} for the ISCX and ISOT datasets, respectively.

While we primarily used a surrogate model to generate adversarial samples, we also tested these samples on the 'best' models to verify the transferability of the adversarial attacks. A direct comparison of the misclassification rates between the surrogate models and the most effective models for each dataset (decision tree for ISCX and Random Forest for ISOT) was needed to verify transferability. Figures \ref{fig:iscx-l2_C&W},\ref{fig:iscx-l2_gan}, \ref{fig:isot-l2_C&W} and \ref{fig:isot-l2_gan}, show the misclassification rate of the surrogate model on the adversarial sample generated and the misclassification of the best model, providing insights on how transferable the attacks are.

\subsection{Experimental setup for Adversarial Retraining and Conformal Prediction}

In our study, after applying the C\&W and GAN adversarial attack on the ISCX and ISOT datasets, we collected a significant number of adversarial samples from 5,100,750,1000,2000 iterations, amounting to a total of 874,036  samples while using C\&W and 1,072,827 while using GAN in ISCX dataset whereas for ISOT dataset 60,5915 while using C\&W samples and 714,006 while using GAN. This structured approach allowed us to incorporate varying levels of adversarial samples. Given the vulnerability observed in the Decision Tree and Random Forrest Model – which otherwise was the best performer for the ISCX and ISOT datasets- we retrain both models. We began by integrating these adversarial samples into the original training set. However, rather than using the same model parameters, we took an extra step to re-optimize the hyperparameters. For this, we employed a genetic algorithm (GA). Using GA ensures that our model is not just learning from the adversarial samples but also being tuned in the best possible manner to accommodate this new, augmented dataset and the best hyperparameter found is shown in Table \ref{tab:retrain_test_set}. Upon retraining with the enhanced dataset and the optimized hyperparameters, we again tested the model against the test sample and the result is shown in \ref{tab:retrain_test_set}. We also make sure the adversarial samples are correctly identified, and the result is shown in Table \ref{tab:retrain-result}.

\begin{figure*}[h]
    \centering
    \begin{minipage}{0.5\textwidth}
        \centering
        \includegraphics[scale=0.35]{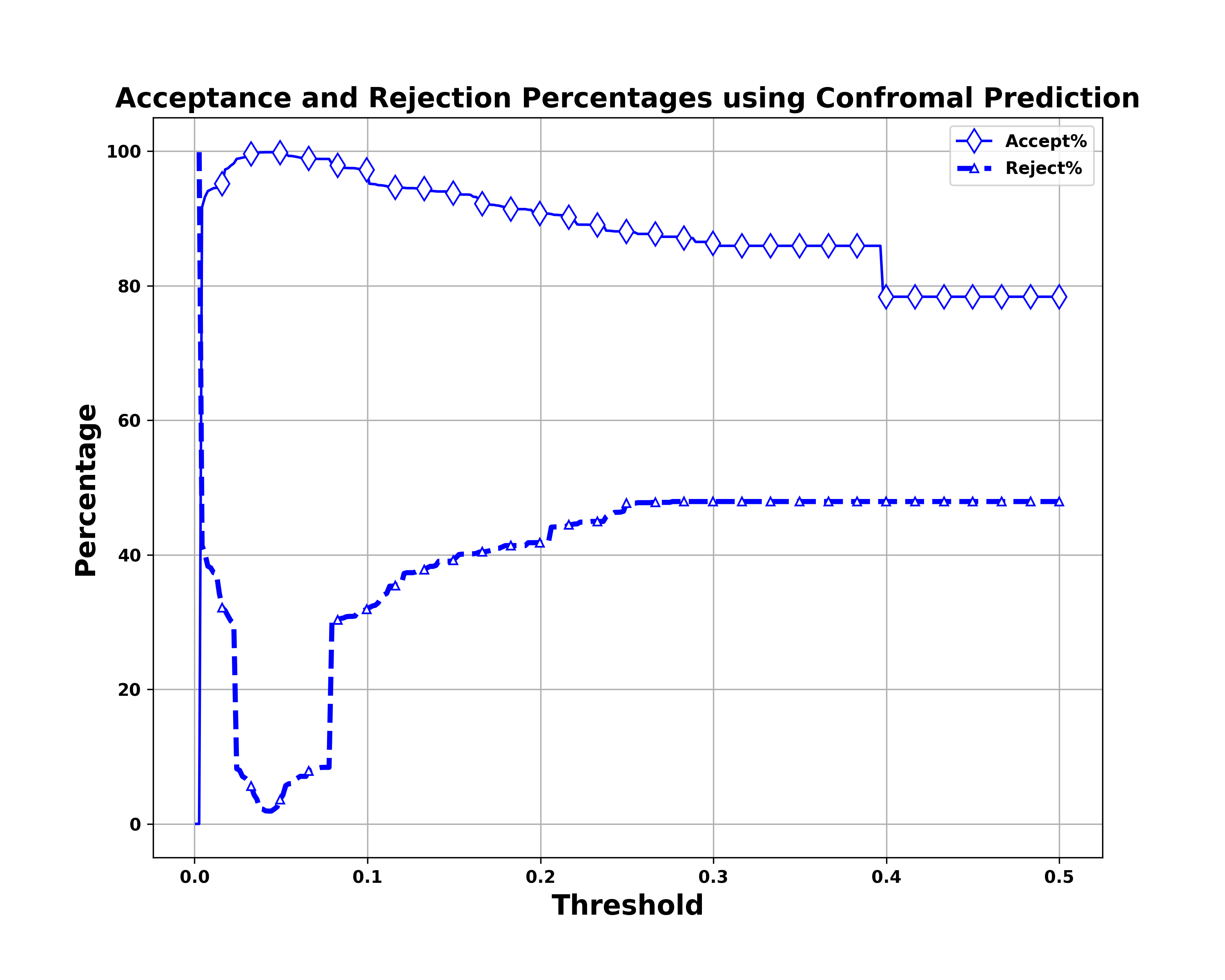}
    \end{minipage}
    \begin{minipage}{0.5\textwidth}
        \centering
        \includegraphics[scale=0.35]{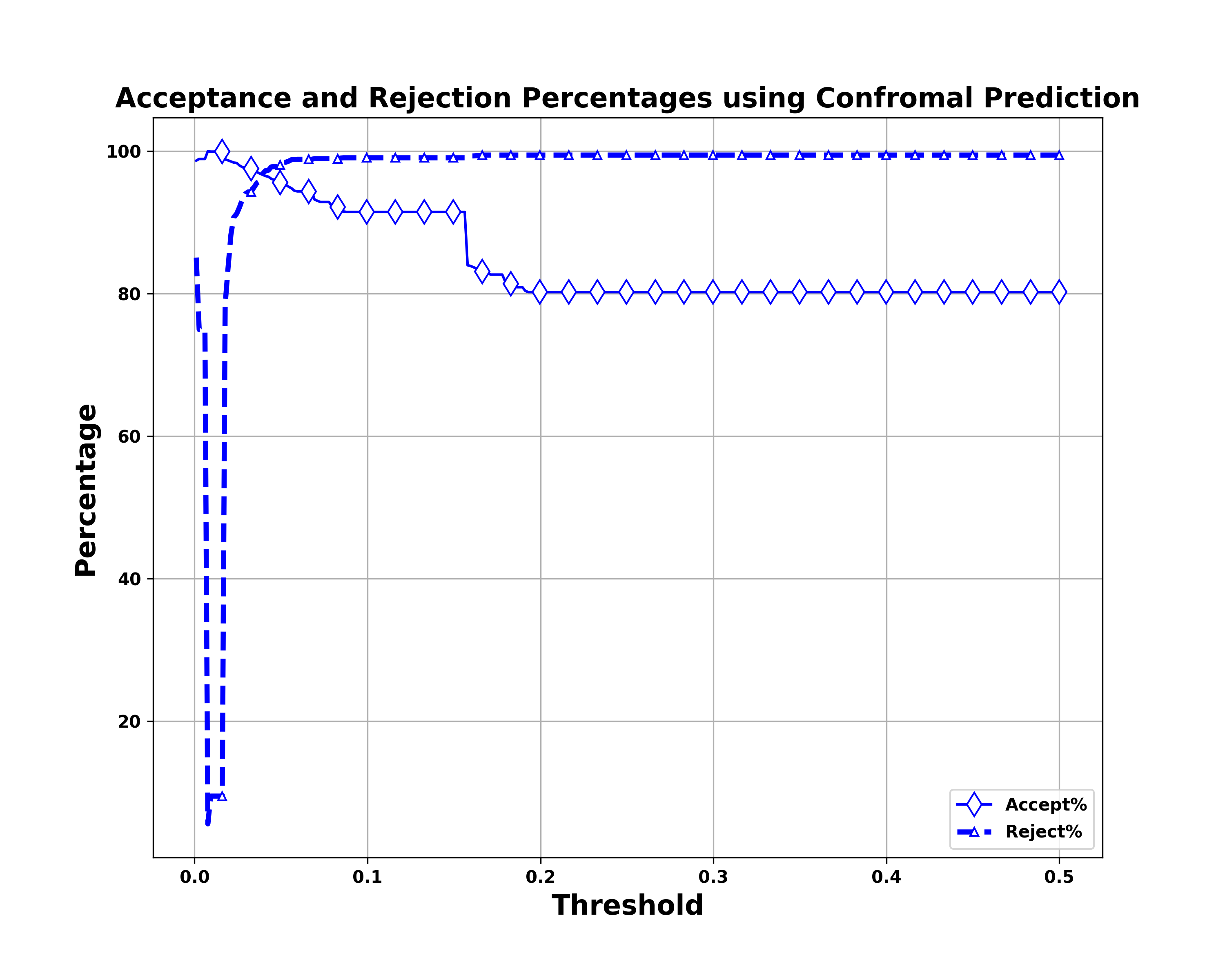}
    \end{minipage}
    \caption{Correctly Accepted and Correctly Rejected Trade-off Graph using a range of $\alpha$  threshold for conformal prediction on ISCX(left) and ISOT(right)}
    \label{fig:datasets-threshold}
\end{figure*}

\begin{table}[!ht] 
\centering
\footnotesize
\caption{{Best hyperparameters using GA optimization on ISCX and ISOT dataset with adversarial sample included in training}}
\begin{tabular}{|l|l|p{8cm}|l|}
\hline
\textbf{Dataset} & \textbf{Classifier} & \textbf{Details} & \textbf{Type} \\ \hline
ISCX & DT & criterion: gini, splitter: best, max\_depth: 18, min\_samples\_split: 11, min\_samples\_leaf: 3, min\_weight\_fraction\_leaf: 0, max\_features: auto, max\_leaf\_nodes: None, min\_impurity\_decrease: 0.0, ccp\_alpha: 0.0 & Hyperparameters \\ \cline{2-4}
 & DT & Accuracy: 90.93\%, Precision: 89.84\%, Recall: 96.03\%, F1 score: 92.83\% & Performance Metrics \\ \hline
ISOT & RF & n\_estimators: 200, criterion: entropy, max\_depth: 42, min\_samples\_split: 2, min\_samples\_leaf: 1, min\_weight\_fraction\_leaf: 0, max\_features: 7, max\_leaf\_nodes: None, min\_impurity\_decrease: 0.0, bootstrap: True, ccp\_alpha: 0.0 & Hyperparameters \\ \cline{2-4}
 & RF & Accuracy: 99.23\%, Precision: 97.97\%, Recall: 91.91\%, F1 score: 94.85\% & Performance Metrics \\ \hline
\end{tabular}
\label{tab:retrain_test_set}
\end{table}

\begin{table}[!ht]
\centering
\footnotesize
\caption{{Adversarial Samples Classifications Before and After Retraining}}
\begin{tabular}{|l|r|r|}
\hline
& \multicolumn{2}{c|}{\bf Before Retraining} \\ \hline
\textbf{Dataset} & \textbf{Benign} & \textbf{Malware} \\ \hline
ISCX & GAN:1,072,827, C\&W:874,036, Total:1,946,863 & 0 \\ \hline
ISOT & GAN:714,006, C\&W:605,916, Total:1,319,922 & 0 \\ \hline
& \multicolumn{2}{c|}{\bf After Retraining} \\ \hline
\textbf{Dataset} & \textbf{Benign} & \textbf{Malware} \\ \hline
ISCX & 121 & 1,946,742 \\ \hline
ISOT & 7 & 1,319,915 \\ \hline
\end{tabular}
\label{tab:retrain-result}
\end{table}





\begin{table}[!htb]
\centering
\caption{Conformal Prediction Performance for ISCX and ISOT}
\label{tab:conformal}
\footnotesize
\begin{tabular}{|l|l|l|l|l|}
\hline
\textbf{Type} & \textbf{Instances} & \textbf{Accept} & \textbf{Accept(\%)} & \textbf{Reject(\%)} \\
\hline
\multicolumn{5}{|c|}{\textbf{ISCX - DT+Conformal}} \\
\hline
\multicolumn{5}{|c|}{\textbf{Correctly Predicted}} \\
\hline
Benign & 85953 & 71429 & 83.10 & 16.90 \\
\hline
Malware & 156808 & 150154 & 95.76 & 4.24 \\
\hline
Total & 242761 & 221583 & 91.28 & 8.72 \\
\hline
\multicolumn{5}{|c|}{\textbf{Incorrectly Predicted}} \\
\hline
Benign & 17726 & 11053 & 62.35 & 37.65 \\
\hline
Malware & 6470 & 3028 & 46.80 & 53.20 \\
\hline
Total & 24196 & 14081 & 58.20 & 41.80 \\
\hline
\multicolumn{5}{|c|}{\textbf{ISOT - RF+Conformal}} \\
\hline
\multicolumn{5}{|c|}{\textbf{Correctly Predicted}} \\
\hline
Benign & 407482 & 387035 & 94.98 & 5.02 \\
\hline
Malware & 31184 & 21324 & 68.38 & 31.62 \\
\hline
Total & 438666 & 408359 & 93.09 & 6.91 \\
\hline
\multicolumn{5}{|c|}{\textbf{Incorrectly Predicted}} \\
\hline
Benign & 643 & 21 & 3.27 & 96.73 \\
\hline
Malware & 2742 & 15 & 0.55 & 99.45 \\
\hline
Total & 3385 & 36 & 1.06 & 98.94 \\
\hline
\end{tabular}
\end{table}

To evaluate the performance and robustness of conformal prediction in network flow, we allocated 10\% of our training data to serve as the calibration set for the conformal prediction process. Utilizing the calibration set, we calculated the conformal scores $s_i$. With these scores, we were able to determine the threshold $\hat{q}$ which is the empirical quantile of the scores from $s_1$ through $s_n$ given by $\lceil(n+1)(1-\alpha)\rceil/n$. To identify the optimal threshold $\alpha$ for accepting or rejecting predictions, we iterate over a spectrum from conservative (high $\hat{q}$) 0.5 to permissive (low $\hat{q}$) 0.001 by generating 200 equally spaced points between the two value and Figure \ref{fig:datasets-threshold} shows the Correctly accepted and Correctly rejected percentage obtained. To maximize the correctly accepted percentage and correctly rejected percentage, we maximize the harmonic mean of the Correctly Accepted Percentage and Correctly Rejected Percentage given in equation \ref{eqn:harmonic_mean}. The best thresholds found are 0.076 in the ISOT dataset and 0.17 in the ISCX dataset. In our conformal prediction framework, an instance would either be accepted if the prediction set is non-empty, indicating a reliable classification or rejected if the prediction set is empty, highlighting uncertainty and necessitating further inspection. We carefully recorded and analyzed the instances categorized as accepted or rejected, simultaneously distinguishing between those correctly and incorrectly predicted. The results, presented in Table \ref{tab:conformal}, offer a comprehensive breakdown of these categories: correctly predicted and accepted, correctly predicted and rejected, incorrectly predicted and accepted, and incorrectly predicted and rejected.

\begin{equation}\label{eqn:harmonic_mean}
  argmax(\alpha) = 2 \times \frac{CA\%(\theta) \times CR\%(\theta)}{CA\%(\theta) + CR\%(\theta)}  
\end{equation}

In this formulation, \(CA\%(\theta)\) represents the percentage of Correctly Accepted predictions, calculated as $\frac{CA(\theta)}{\sum{1}\{y_i = \hat{y}_i\}} \times 100$, while \(CR\%(\theta)\) signifies the percentage of Correctly Rejected predictions, computed as $\frac{CR(\theta)}{\sum {1}\{y_i \neq \hat{y}_i\}} \times 100$.

\subsection{Experimental Analysis}
\subsubsection{Evaluating Feature Sensitivity: Average L2 distance and Misclassification Rates}

In order to comprehensively evaluate the subtlety and effectiveness of adversarial perturbations, we focused on two key metrics: the change in average L2 distances between original and adversarial samples and the misclassification rate of the surrogate model. The average L2 distance offers insights into the magnitude of perturbations; lower values signify more imperceptible changes, while higher values may hint at overt alterations. On the other hand, the misclassification rate gauges the success of these perturbations in misleading the model.

\textbf{For the ISCX dataset},the baseline misclassification rate of the 
surrogate model was identified as 9.45\%. During the C\&W attack, Starting with the 'SrcWin' feature, a gradual increase in misclassification rates was observed, moving from the baseline of 9.45\% and stabilizing at 60\%. This uptrend coincided with an increase in the L2 distance, suggesting a direct correlation between the magnitude of perturbation and the likelihood of misclassification. However, after 250 iterations, the generation of adversarial samples ceased, implying a threshold of robustness specific to the 'SrcWin' feature. The 'Shops' feature initially presented a misclassification rate slightly above the baseline at 14\%, with an L2 distance starting at 191.0. As iterations progressed, the L2 distance increased marginally to 191.6, yet the misclassification rate intriguingly converged towards the baseline, settling at 9.5\%. This pattern indicates the model's capacity to adapt to and mitigate the impact of increasing perturbations on the 'Shops' feature. Conversely, the 'sTtl' and 'dTtl' features exhibited a consistent vulnerability, with misclassification rates persistently at 100\%. The L2 distance for these features showed a slight decline, suggesting that even minimal perturbations were sufficient to compromise the model's accuracy, highlighting a significant susceptibility in handling TTL-related attributes. In examining the 'SrcBytes' and 'DstBytes' features, a notable decrease in L2 distance was observed, from 650,000 to 450,000 and from 800,000 to 450,000, respectively. Despite the reduction in perturbation magnitude, the misclassification rates remained around 20\%, showing the efficacy of the attack in deceiving the model even with subtler alterations. The 'Duration' feature demonstrated an initial decrease in L2 distance, indicating an attempt to refine the adversarial perturbations. However, the misclassification rate showed minimal variation, suggesting that the model's performance could be influenced by more pronounced perturbations, as evidenced by the subsequent increase in L2 distance.
'TotBytes' and 'Rate' features further elucidated the model's response spectrum to adversarial challenges. 'TotBytes' experienced a substantial reduction in L2 distance, which correlated with an improvement in classification accuracy, as the misclassification rate declined to 17.5\%. Meanwhile, the 'Rate' feature underwent a significant drop in L2 distance but only saw a slight decrease in the misclassification rate, suggesting that certain features may require more nuanced adjustments to enhance the model's resilience. In summary, 'SrcWin' and 'Shops' showcase the model's capacity to withstand increasing perturbations; others, notably TTL-related features, highlight intrinsic vulnerabilities that adversarial attacks can exploit. 

During the GAN attack of the ISCX dataset, the 'SrcWin' feature experienced an incremental rise in L2 distance, from an ideal baseline of zero to 2.33, which corresponded with an elevation in the misclassification rate from 30\% to a stable 60\%. This trend indicates a saturation point in the classifier's vulnerability, beyond which additional perturbations ceased to increase the misclassification rate, suggesting a nuanced robustness of the classifier against adversarial manipulations targeting the 'SrcWin' feature. Conversely, the 'Shops' feature showcased a different pattern, where the L2 distance started at approximately 20 but decreased gradually to around 3 and with it, the misclassification decreased linearly. Significant findings were observed in the TTL-related features, 'sTtl' and 'dTtl', where the model exhibited a consistent vulnerability. For 'sTtl', the L2 distance escalated from 35 to 58, accompanying a surge in misclassification rate to a consistent 100\%. Similarly, 'dTtl' maintained a 100\% misclassification rate throughout, with the L2 distance showing a minor ascent from 34.5 to 34.1. These outcomes emphasize the classifier's susceptibility to adversarial attacks on TTL attributes, where even slight perturbations significantly compromise classification accuracy. For 'SrcBytes', a noteworthy reduction in L2 distance from 25,000 to 0 was observed, and with it, the misclassification decreased from 14 to 11. A similar pattern emerged for 'DstBytes', where the L2 distance saw a substantial reduction from 80,000 to 0, yet the misclassification rate persistently hovered around 20\%. These instances demonstrate the classifier's consistent challenge in accurately discerning adversarial samples from genuine ones, even as perturbations became less pronounced. The 'Duration' feature revealed an initial decline in L2 distance from 20 to approximately 0, which was met with a decrease in misclassification rate from 40\% to 11\%. Similar patterns were observed in  TotBytes' features. Conversely, the 'Rate' feature didn't change its L2 distance much, staying at around 2100, and the misclassification remains consistently around 100\%. 

Upon analyzing the ISCX dataset subjected to both C\&W and GAN adversarial attacks, it's evident that each method exhibits distinct characteristics in terms of perturbation efficiency and the resulting misclassification rates across various network features. The nuanced examination reveals that the GAN attack generally produces adversarial samples with lower L2 distances compared to those generated by the C\&W attack, suggesting a subtler approach to manipulating the data while still effectively deceiving the classifier. For instance, during the GAN attack, features like 'SrcBytes' and 'Duration' saw a notable decrease in L2 distance, approaching nearly zero, which did not significantly compromise the model's ability to classify correctly, as seen in the modest decrease in misclassification rates. This contrasts with the C\&W attack, where, despite achieving high misclassification rates for features like 'sTtl' and 'dTtl', the approach necessitated larger perturbations, as indicated by the relatively higher L2 distances. Such differences highlight the GAN attack's capability to exploit the model's vulnerabilities with minimal deviation from the original dataset, making these perturbations less detectable and potentially more dangerous. Moreover, the effectiveness of the GAN attack in maintaining or even lowering L2 distances while achieving high misclassification rates, particularly for features such as 'SrcWin' and 'Shops', shows its effectiveness. The 'SrcWin' feature, under the GAN attack, exhibited an increase in misclassification rates to 60\% with a moderate increase in L2 distance to 2.33, showcasing the attack's precision in targeting the model's weaknesses. Conversely, the C\&W attack, though effective in increasing misclassification rates, often required more significant alterations to the data, as seen in the increased L2 distances for the same and other features like 'Shops'. Interestingly, TTL-related features ('sTtl' and 'dTtl') consistently showed vulnerability under both attacks, with 100\% misclassification rates, showing an area for enhancing the classifier's defences.

 \textbf{In the ISOT dataset}, the baseline misclassification rate of the surrogate model was identified as 13.87\%. During the C\&W attack, starting with the 'SrcWin' feature, we observed an initial average L2 distance of approximately $1.75 \times 10^3$, which corresponded with a misclassification rate of 32.5\%. Over the iterations, this L2 distance narrowed closer to the original data, culminating in a reduced misclassification rate of about 15\% by the 750th iteration. This trajectory suggests that the model becomes more adept at identifying adversarial samples as they become subtler, indicating a growing challenge for the attack to deceive the model effectively. The 'Shops' feature exhibited a contrasting scenario with an astoundingly high initial misclassification rate of 99\%, despite only a minimal increase in the average L2 distance from $1.2708 \times 10^2$ to $1.2713 \times 10^2$. 'sTtl' and 'dTtl' observe similar patterns with Shops where the misclassification rate is very high with little change in the Initial L2 distance. 'SrcBytes' saw an L2 difference starting at $1 \times 10^{13}$, halved over the iterations, accompanied by a fluctuating misclassification rate that peaked at 36\% during the 750th iteration before dropping to 22\%. This indicates a diminishing deception efficiency as adversarial examples more closely resemble genuine data. A similar trend was noted for 'DstBytes', with the L2 difference commencing at $1.4 \times 10^{12}$ and reducing to $7 \times 10^{11}$, alongside a temporary spike in misclassification to 82\%, which then slightly reduced to 79.5\% by the end of the iterations. The 'Duration' feature, despite a significant reduction in L2 difference from $1.2 \times 10^3$ to $2 \times 10^2$, maintained a relatively stable misclassification rate around 14.1\%, suggesting the model's inherent robustness to perturbations affecting this feature. The 'TotBytes' feature showed a nearly consistent misclassification rate of around 88\% with the L2 distance fluctuating around $1 .7\times 10^{112}$ to $9 \times 10^{11}$. Finally, the Rate feature shows a decrease in misclassification rate of 100\% to 90\% with its L2 distance fluctuating between $1 .4\times 10^{08}$ to $8 \times 10^{07}$. In summary, features like 'Shops', 'sTtl', and 'dTtl' remain significantly vulnerable despite minimal perturbation changes. The consistent misclassification rate for the 'Duration' feature despite varying L2 distances underscores the feature resilience against adversarial perturbations.
 
During the GAN attack of the ISOT dataset, The 'SrcWin' feature showcased a remarkable improvement in the classifier's ability to discern between genuine and adversarial samples, as evidenced by the reduction in L2 distance from approximately 3.50 to near zero, accompanied by a decrease in misclassification rate from around 87\% to 10\%. This significant improvement indicates that the neural network became increasingly effective at classifying samples correctly as adversarial perturbations were refined to more closely resemble the original data. In stark contrast, the 'Shops' feature maintained a stubbornly high misclassification rate of 100\%, despite a halving of the L2 distance from around 3 to 1.5. This persistence in high misclassification rates, regardless of a reduction in perturbation magnitude, highlights a particular vulnerability of the model to attacks targeting the 'Shops' feature, suggesting an area where the model's defences could be bolstered. The TTL-related features 'sTtl' and 'dTtl' presented a uniform challenge, with both features experiencing a consistent 100\% misclassification rate. For 'sTtl', the L2 distance increased from about 40 to 65, and for 'dTtl', it showed a negligible rise from 39.68 to 39.70, underlining the model's ongoing susceptibility to adversarial manipulations affecting TTL attributes despite varying degrees of perturbation. The 'SrcBytes' feature, with a substantial reduction in L2 distance from $1x10^12$ to nearly 0, saw a corresponding decrease in misclassification from approximately 85\% to 15\%. Similarly, the 'Duration' feature witnessed a decrease in L2 distance from around 3.0 to nearly 0, with misclassification rates dropping from about 80\% to 15\%, demonstrating the model's enhanced performance as adversarial examples became increasingly indistinguishable from genuine data. However, the 'DstBytes' and 'TotBytes' features displayed 100\% with a very large L2 distance. Lastly, the 'Rate' feature underscored a direct correlation between increased perturbation magnitude and heightened vulnerability, with L2 distance expanding from 45 to 250 and misclassification rates climbing from 70\% to 100\%. In summary, While certain features like 'SrcWin' and 'Duration' evidenced the model's potential for improvement and resilience, others, notably 'Shops,' 'sit,' 'dTtl,' 'DstBytes,' and 'TotBytes,' underscored persistent vulnerabilities.

In comparing C\&W and GAN attacks on the ISOT dataset, certain features like 'Shops', 'sTtl', 'dTtl', and 'DstBytes' displayed a consistent vulnerability under both attack types, with high misclassification rates that were relatively unaffected by the scale of L2 distance changes. This indicates that these features are inherently more susceptible to adversarial manipulations, regardless of the attack method. While the C\&W attack demonstrates the potential for inducing high misclassification rates with larger perturbations, the GAN attack distinguishes itself by achieving similar or even higher rates of misclassification with subtler, less detectable changes in L2 distance.

\subsubsection{Experimental Analysis of Adversarial Attack Transferability}

To understand the transferability of the attack in the ISCX dataset, we tested adversarial examples that are successful against the surrogate model on a Decision Tree. The original Decision Tree had a misclassification rate of 3.42\%. For features like 'SrcWin' and 'Duration', the Decision Tree held its ground with misclassification rates near this baseline for the C\&W attack; however, during the GAN attack, the misclassification increased to 35\%. For 'Shops', the rate increased to 7\% before settling back to 3.4\% during the C\&W attack and during the GAN attack, the misclassification increased to around 85\% but eventually dropped down to around 15\% as the L2 distance became smaller noticeably, the surrogate model, and the Decision Tree displayed same misclassification rate implying the attacks are transferable. The 'sTtl' and 'dTtl' features significantly rose, with rates hovering at around 50\% during both attacks, indicating the tree's vulnerability for these features. In contrast, 'TotBytes' and 'Rate' only saw minor increases in both attacks, with rates around 4\%. This highlighted that while the surrogate model (NN) was susceptible to the adversarial attack, the Decision Tree largely resisted the attack on most of the features except for sTtl and dTtl in both attacks. Although SrcWin and SHops are hard to attack using C\&W attack, with GAN, the two features saw an increase in the misclassification rate.

For the ISOT dataset, in the 'SrcWin' feature, the attack pushed the RF model's misclassification rate up to 20\% for C\&W and 60\% for GAN. However, this heightened misclassification dipped back to the 7.7\% baseline as we iterated. This swing back to the baseline suggests the C\&W attack might lose some of its edge as its perturbed samples draw closer to genuine data. The Shops feature consistently gets a misclassification rate of 93\% during the C\&W attack and 94\% during the GAN attack. In the 'sTtl' feature, GAN achieves a better misclassification rate as compared to the C\&W attack, staying constant at 90\% for GAN, but for C\&W, the misclassification drops down to 40\%. For 'dTtl', it converged to around 35\% for the C\&W attack, but for GAN, the misclassification drops down from 90\% to around 21\% as the iteration progresses. The 'SrcBytes' feature saw its highest misclassification around the 750th iteration, after which it decreased to 15\% during the C\&W attack, whereas during GAN, the misclassification decreased from 60\% to around 15\% as the iteration progressed.
During the C\&W attack of the 'DstBytes' misclassification rate increased at 70\% around the same 750th iteration mark and settled at 68\% as the iteration progressed; however, for GAN, the misclassification remained constant at around 75\%. Interestingly, the 'Duration' feature remained largely unfazed by the C\&W attack, holding its misclassification rate steady at 7.7\%; however, during the GAN attack, the misclassification drops down from 60\% to the baseline as L2 distance decreases. The 'TotBytes' feature experienced a relatively similar misclassification rate of around 75\% during both attacks. Finally, 'Rate' misclassification started at 81\% and was reduced to 74\% during the C\&W attack, whereas the misclassification rate increased from 20\% to around 80\% for the GAN attack.

Our comprehensive examination across the ISCX and ISOT datasets has shown the intricate nature of adversarial sample transferability and its varying impact on classifier performance. Specific features such as 'sTtl', 'dTtl', and 'Shops' demonstrated a heightened vulnerability, with adversarial samples successfully compromising the model's integrity. Contrastingly, features like 'Duration' showcased a commendable resilience against C\&W; however, GAN shows the feature is still vulnerable to its attack.
Crucially, it is observed that the misclassification rates induced by adversarial samples are consistently lower when tested on the primary models (Decision Tree for ISCX and Random Forest for ISOT) in comparison to the surrogate Neural Network model. Confirming not all the samples generated by the surrogate model is transferable.

\subsubsection{Experimental Analysis of Adversarial Retraining and Conformal Prediction}

Upon the completion of the adversarial retraining process, the models displayed remarkable improvement in their capability to correctly identify adversarial samples, as can be seen in Table \ref{tab:retrain-result}. In the initial state, prior to retraining, the models were unable to correctly classify any of the adversarial samples in both the ISCX and ISOT datasets, misclassifying them all as benign, which changed post-retraining. The Decision Tree model misclassified a mere 121 out of the 1,946,863 adversarial samples in the ISCX dataset, showcasing a significant leap in its ability to detect and correctly classify adversarial inputs. In the case of the ISOT dataset, the Random Forest model demonstrated even more impressive performance, only misclassifying 7 adversarial samples. In addition to this notable improvement in adversarial sample detection, the models also maintained robust performance metrics on the test set, as shown in Table \ref{tab:retrain_test_set}.

Conformal prediction has played an indispensable role in enhancing the trustworthiness and reliability of our models, as demonstrated by the comprehensive results laid out in Table \ref{tab:conformal}. By implementing this technique, we have been able to substantially increase our confidence in the model's predictions, thus ensuring that the outputs are not only accurate but also dependable. In the context of the ISCX dataset, when integrated with the Decision Tree model, conformal prediction exhibited exceptional performance. A staggering 91.28\% of the total instances were correctly accepted, signifying that the model, when combined with conformal prediction, is highly adept at recognizing and affirming correct predictions. This is a crucial aspect, especially in network security applications, where the cost of false negatives can be substantial. On the flip side, the model also showcased its proficiency in identifying potential misclassifications, successfully rejecting 58.20\% of the incorrectly predicted instances. This indicates that the model is not only good at affirming what it knows but also at flagging what it does not, adding a layer of security and reliability. The effectiveness of conformal prediction was further shown in the ISOT dataset, where it was employed alongside the Random Forest model. Here, it correctly accepted 93.09\% of the total instances, showcasing its ability to discern and uphold accurate predictions. More impressively, it rejected an overwhelming 98.94\% of the incorrect predictions. This high rejection rate is particularly noteworthy, as it highlights the model's capacity to effectively eliminate unreliable outputs, ensuring that the predictions made are both precise and dependable.

By striking a delicate balance between accepting correct predictions and rejecting incorrect ones, conformal prediction has proved to be an invaluable asset in the classification process. It has not just bolstered the accuracy of our models but has also added a level of reliability and robustness that is paramount in network security contexts. This dual capability ensures that our models are not just performing well but also providing outputs that we can trust, making them indispensable tools in network security.

\subsection{Performance Comparison with Other Works}

\begin{table}[!ht]

\centering
\footnotesize
\caption{{Comparison of Related Works on ISCX Dataset}}
\begin{tabular}{|p{5cm}|p{2.1cm}|p{2.5cm}|}
\hline

\textbf{Author} & \textbf{ Classifier} & \textbf{Performance} \\ \hline
Li, Y. \& Yao 2022 \cite{li2022botnet}& ResNet & Acc=93.67\%, Prec=92.2\%, F1=93.0\% \\ \hline
Li, Y. \& Yao 2022 \cite{li2022botnet}  & CBAM-Resnet & Acc=95.85\%, Prec=95.26\%, F1=95.73\% \\ \hline 
Hassan et al 2021\cite{hassan2021intrusion}. & Payload Embedding & Acc=80.9\%, Prec=80.9\%, Rec=80.9\%, F1=80.9\%\\
\hline

Li, Y. \& Yao 2022 \cite{li2022botnet}  & Parallel CBAM-ResNet and Self-attention & Acc=97.26\%, Prec=96.94\%, Rec=97.15\%\\ \hline
Shahhosseini et al. 2022 \cite{shahhosseini2022deep} & LSTM & Prec=95\%,  Rec=94\%,  F1=94.5\%\\ \hline
Guangli 2024 \cite{wu2024bot} & Bot-DM: Multilayer Transformer and DNN & Acc=91.92\%, Prec=91.45\%, Rec=91.45\%\\ \hline 

Meher Afroz 2024 \cite{afroz2024feature} & DT & Acc=81.71\%, Prec=83.92\%\\ \hline

*Our work (without Adversarial Hardening and Conformal Prediction) &  DT   & Acc=93.83\%, Prec=93.55\%, Rec=96.57\%, F1=95.04\% \\ \hline
*Our work (with Adversarial Hardening and Conformal Prediction) & DT   & Acc=94.02\%, Prec=93.14\%, Rec=98.02\%, F1=95.52\% \\ \hline
\end{tabular}
\begin{flushleft} Legend: Acc-Accuracy, Prec-Precision, Rec-Recall, F1-F1-score
\end{flushleft}
\label{tab:comparison_iscx}

\end{table}

\begin{table}[!ht]

\centering
\footnotesize
\caption{{Comparison of Related Works on ISOT Dataset}}

\begin{tabular}{|p{5cm}|p{3cm}|p{3cm}|}
\hline

\textbf{Author} & \textbf{ Classifier} & \textbf{Performance} \\ \hline
Mai \& Park 2016 \cite{Mai} & K-means & Detection=97.11\% \\ \hline
Pektas \& Acarman 2017 \cite{pektacs2017effective} & RF  & F1=99.0\%, Acc=99.5\%, Rec=99.0\%, Prec=99.0\% \\ \hline

Debashi, M et al 2018\cite{{debashi2018sonification}}& SoNSTAR & Acc=99.2\%, Prec=97.1\%, Rec=99.5\%, F1=98.3\%,  \\\hline 
Khan et al. 2019 \cite{Khan} & DT & Acc=98.7\% \\ \hline

Mehdi Asadi et al 2020\cite{asadi2020detecting} & DNNSVMLib-c4.5 & Acc=99.64\% \\ \hline

*Our work (without Adversarial Hardening and Conformal Prediction) &  RF  & Acc=99.27\%,  Prec=98.16\%,  Rec=92.3\%, F1=95.14\% \\ \hline
*Our work (with Adversarial Hardening and Conformal Prediction) &  RF  & Acc=99.99\%, Prec=99.90\%, Rec=99.93\%, F1=99.92\% \\ \hline

\end{tabular}
\begin{flushleft} Legend: Acc-Accuracy, Prec-Precision, Rec-Recall, F1-F1-score
\end{flushleft}
\label{tab:comparison_isot}

\end{table}
In our research, we have primarily focused on identifying adversarial samples, assessing feature vulnerabilities, and analyzing the transferability of adversarial samples. Despite this, we have also evaluated the performance of classifiers in our method and compared them with other research papers.

In the ISCX dataset comparison shown in Table \ref{tab:comparison_iscx}, the Decision Tree (DT) classifier shows competitive results compared to single classifiers from other works. Without the conformal prediction layer and adversarial retraining, our DT classifier achieves an F1 score of 95.04\%, which is comparable to the results from Li, Y. \& Yao (2022) \cite{li2022botnet} who used ResNet and CBAM-Resnet, achieving F1 scores of 93.0\% and 95.73\%, respectively. However, when we integrate the conformal prediction layer, our DT classifier's performance is enhanced significantly, reaching an F1 score of 95.52\%. This surpasses the single classifiers' performance and is comparable to the Parallel CBAM-ResNet ensemble and Self-attention from Li, Y.\& Yao (2022), which achieved an F1 score of 97.15\%. The LSTM from Shah Hosseini et al. (2022) \cite{shahhosseini2022deep} achieved an F1 score of 94.5\%, and the ensemble approach of LSTM-RF has a better result. However, it is crucial to highlight that these methods utilize multiple classifiers, and integrating a conformal prediction layer on top of these ensemble methods could yield even better performance. Guangali\cite{wu2024bot} utilize a multilayer Transformer and DNN, achieving an accuracy of 91.92\%, which is comparable to the accuracy obtained by our work 95.04\% which can be further improved if conformal layers are implemented along with the model. Meher Afroz\cite{asadi2020detecting} utilize DT, but there is no hyperparameter optimization, which likely impacts their performance.

In the ISOT dataset comparison Table \ref{tab:comparison_isot}, our method with the Random Forest (RF) classifier and conformal prediction layer outperforms all other methods listed, achieving an F1 score of 99.92\%. This is a significant improvement over other techniques, such as SoNSTAR by Debashi et al. (2018) \cite{debashi2018sonification}, which achieved an F1 score of 98.3\%, and the DT classifier by Khan et al. (2019) \cite{Khan}, which had an accuracy of 98.7\%. Mehdi et al. \cite{asadi2020detecting} that using DNNSVMLib achieved a similar result with an accuracy of 99.64\%, which is slightly lower than our accuracy of 99.99\%, which might be just statistical noise, but they didn't provide other important metrics such as Precision, Recall and F1.

In addressing the challenges posed by adversarial attacks in network intrusion detection, our research introduces a novel analytical perspective by rigorously examining the vulnerability of model features through an in-depth analysis of the L2 distance and iteration counts required for successful misclassification. Unlike existing studies, such as those employing standard GAN and C\&W attack \cite{alhajjar2021adversarial, han2021evaluating, chen2020generating}, our approach provides a feature-wise evaluation of feature sensitivity under adversarial conditions. By comparing the efficacy of GAN and C\&W attacks, we demonstrate that GANs are capable of achieving misclassification at significantly lower L2 distances, thus highlighting their potential for more efficient adversarial strategies. Additionally, our work explores the Discriminatory aspect of transferability in adversarial attacks, an area often overlooked in other studies. We investigate the application of adversarial examples across different models, revealing that while some attack samples deceive a surrogate model, they do not always generalize across different feature sets due to varying misclassification rates. This insight is visually supported by Figures 4 and 5, which illustrate the nuanced relationship between perturbed features' L2 distances and their corresponding classification rates of the surrogate model (Neural Network) and the actual model based on a tree-based algorithm across different datasets.

In a significant advancement over traditional methods, our study introduces the application of a conformal prediction layer to network intrusion detection systems (NIDS), focusing on mitigating the impact of uncertain predictions. Unlike other approaches that primarily aim to enhance overall model accuracy, our method leverages the robust framework of conformal prediction to specifically reject uncertain outcomes, thereby improving the reliability and trustworthiness of the predictions. This technique, which can be universally applied across any classifier, has demonstrated substantial improvements in model performance within our tests. Specifically, implementing the conformal prediction layer increased the F1 score from 93.75\% to 95.32\% on the ISCX dataset and from 95.12\% to 99.79\% on the ISOT dataset. 

Our method ensures that the classifiers are high-performing in accuracy, precision, recall, and F1 score and resilient against adversarial manipulations. This dual focus provides a more comprehensive and reliable solution for network security. Our method, therefore, offers a balanced approach that does not sacrifice the ability to identify adversarial samples for high performance on clean data, ensuring robustness in real-world scenarios where adversarial attacks are a constant threat.

\section{Limitations}
The enhancement of F1 scores through the implementation of a conformal prediction layer has been demonstrated; however, it comes with a cost of rejecting correct predictions, and future studies can explore the reduction of rejection of correct predictions. The efficacy of both the classifiers and the conformal prediction layer is heavily contingent upon the quality and representativeness of the training data. In scenarios where the training data are not sufficiently diverse or fail to encapsulate the full spectrum of potential attack vectors and normal traffic patterns, the model's performance may not accurately reflect its effectiveness in real-world applications. Despite advancements aimed at enhancing robustness against adversarial attacks, the dynamic and evolving nature of these threats may still pose challenges. Newly developed or previously unseen adversarial strategies could potentially compromise the models. Ongoing adaptation and refinement of the models, informed by the latest adversarial tactics, are imperative to sustain high-security levels. Our investigation into the transferability of adversarial examples highlights significant complexities and dependencies on specific model characteristics. Although our findings provide valuable insights, they represent only a preliminary exploration within a complex, multifaceted research area. Further studies are needed to elaborate on these dynamics, potentially broadening the investigation to include diverse models and adversarial conditions.

\section{Time Complexity Analysis}
This section delves into the computational complexities associated with various components of our study, providing insights for understanding the computational demands of the employed methodologies.

The primary classifiers under consideration are the Decision Tree (DT) and Random Forest (RF), utilized respectively in the ISCX and ISOT datasets. The Decision Tree exhibits a time complexity of $O(N \cdot M_f \cdot \log(N))$, where $N$ denotes the number of training examples, and $M_f$ represents the number of features. On the other hand, the Random Forest, an ensemble of multiple decision trees, has a time complexity of $O(N \cdot M_f \cdot \log(N) \cdot T_{RF})$, with $T_{RF}$ indicating the number of trees within the forest. 

In hyperparameter optimization, the Genetic Algorithm (GA) was identified as the optimal choice through comparative analysis with Particle Swarm Optimization (PSO). The GA initializes populations with a complexity of $O(P \cdot n_{hp})$, where $P$ stands for the population size and $n_{hp}$ signifies the number of hyperparameters. Each population's evaluation demands $O(P \cdot E)$ time complexity, with $E$ encapsulating the classifier training and fitness function computation. The tournament selection process in GA adds a complexity of $O(P \cdot T_{GA})$, and the crossover operation necessitates $O(P \cdot n_{hp})$ time. Aggregating these complexities, the overall per generation complexity of the GA algorithm sums up to $O(G \cdot P \cdot (3n_{hp} + E + T_{GA}))$, with $G$ representing the number of generations. When considering the Decision Tree classifier inclusive of hyperparameter optimization, the resultant time complexity is $O(N \cdot M_f \cdot \log(N)) + O(G \cdot P \cdot (3n_{hp} + E_{DT} + T_{GA}))$. Similarly, for the Random Forest classifier, the overall time complexity is expressed as $O(N \cdot M_f \cdot \log(N) \cdot T_{RF}) + O(G \cdot P \cdot (3n_{hp} + E_{RF} + T_{GA}))$.

The C\&W attack methodology involves multiple computational steps. The calculation of the L2 norm, with $D$ representing the input space's dimensionality, incurs a time complexity of $O(D)$. Neural network output evaluation, denoted as $Z(x')$, and subsequent operations result in a complexity of $O(M_{NN} + C)$, where $M_{NN}$ encapsulates the neural network evaluation complexity, and $C$ represents the number of classes. The gradient computation through backpropagation, alongside additional operations for the perturbation term, introduces a complexity of $O(D + M_{NN})$. Factoring in the iterative nature of the Adam optimization algorithm, denoted by $I$ iterations, the cumulative time complexity of the C\&W attack approximates to $O(I \cdot (D + M_{NN} + C))$. 

The GAN attack consists of training two neural network Generator and Discriminator. The time complexity of neural network is  $ O(t \cdot n \cdot \sum_{i=1}^{n-1} x_i \cdot x_{i+1}) $.Let the generator \( G \) have \( n_G \) layers with \( g_1, g_2, \ldots, g_{n_G} \) nodes in each respective layer and the discriminator \( D \) has \( n_D \) layers with \( d_1, d_2, \ldots, d_{n_D} \) nodes in each respective layer. If \( t \) is the number of training examples, \( n \) is the number of epochs. Assuming the discriminator is updated \( k \) times for each update of the generator, the total complexity is $O\left(t \cdot n \cdot \left(k \cdot \sum_{i=1}^{n_D-1} d_i \cdot d_{i+1} + \sum_{i=1}^{n_G-1} g_i \cdot g_{i+1}\right)\right) $.

In conformal prediction, the calibration set, consisting of $N_c$ instances, necessitates probability estimates and conformal scores computation, resulting in a time complexity of $O(N_c \cdot M_f)$. The subsequent threshold determination requires sorting the conformal scores, adding a $O(N_c \cdot \log(N_c))$ complexity. For the test set, comprising $N_t$ instances, the formulation of prediction sets demands a time complexity of $O(N_t \cdot D_t)$ for Decision Trees and $O(N_t \cdot T_{RF} \cdot D_{RF})$ for Random Forests. This results in a total time complexity of $O(N_c \cdot M_f + N_c \cdot \log(N_c) + N_t \cdot D_t)$ for Conformal Prediction with Decision Trees and $O(N_c \cdot M_f + N_c \cdot \log(N_c) + N_t \cdot T_{RF} \cdot D_{RF})$ for Random Forests. The final time complexity of the whole process will be $O(N \cdot M_f \cdot \log(N) + G \cdot P \cdot (n_{hp} + E_{DT} + T_{GA}) + N_c \cdot M_f + N_c \cdot \log(N_c) + N_t \cdot D_t + I \cdot (D + M_{NN} + C))$ for Decision tree and $O(N \cdot M_f \cdot \log(N) \cdot T_{RF} + G \cdot P \cdot (n_{hp} + E_{RF} + T_{GA}) + N_c \cdot M_f + N_c \cdot \log(N_c) + N_t \cdot T_{RF} \cdot D_{RF} + I \cdot (D + M_{NN} + C))$ for Random Forrest.

In the case of the Decision Tree (DT) and Random Forest (RF) classifiers, the convergence of the algorithm where the model's performance stabilizes and does not significantly improve with further training is highly influenced by the tree depth, number of trees, and the complexity of the data. For the DT classifier, it generally converges upon sufficiently partitioning the feature space. For RF, convergence is achieved when additional trees do not markedly improve the model's performance. These factors directly impact the time complexities of $O(N \cdot M_f \cdot \log(N))$ and $O(N \cdot M_f \cdot \log(N) \cdot T_{RF})$ for DT and RF respectively, where a faster convergence could potentially lead to reduced computation time.

In hyperparameter optimization using the Genetic Algorithm (GA), convergence occur when the values of the objective function (such as accuracy, F1 score, etc., for a classifier) stop showing significant changes over generations, the convergence rate is contingent on the number of generations $G$, population size $P$, and tournament size $T_{GA}$. These parameters influence the overall time complexity of $O(G \cdot P \cdot (n_{hp} + E + T_{GA}))$. The more the number of $G$ and $P$ larger the exploration space of the hyperparameter.

The C\&W and GAN attack convergence is Discriminatoral, especially given its iterative nature. In C\&W the associated time complexity is $O(I \cdot (D + M_{NN} + C))$ where, $I$ represents the number of iterations required for convergence. Similarly for GAN the associated time complexity is $O\left(t \cdot n \cdot \left(k \cdot \sum_{i=1}^{n_D-1} d_i \cdot d_{i+1} + \sum_{i=1}^{n_G-1} g_i \cdot g_{i+1}\right)\right)$ where $n$ represents the number of iteration. A higher number of iterations might yield a more precise adversarial example but at an increased computational cost, necessitating a careful calibration of $I$ and $n$ to ensure efficiency.

Conformal prediction, though not iterative, still requires careful consideration of the calibration set size $N_c$ and the number of test instances $N_t$, as these parameters influence the time complexities of $O(N_c \cdot M_f + N_c \cdot \log(N_c) + N_t \cdot D_t)$ and $O(N_c \cdot M_f + N_c \cdot \log(N_c) + N_t \cdot T_{RF} \cdot D_{RF})$ for Decision Trees and Random Forests, respectively.

\section{Conclusion}
Our research has made significant strides in advancing network security defences, particularly in the realm of botnet detection and adversarial sample mitigation. By leveraging both machine learning and deep learning algorithms and fine-tuning their hyperparameters with Genetic Algorithms and Particle Swarm Optimization, we established a strong foundation and achieved optimal predictive accuracy. Our in-depth analysis of feature vulnerabilities using GAN and C\&W attack method revealed crucial insights, allowing us to maintain meaningful semantic and syntactic relationships even when features were manipulated. This meticulous approach to adversarial example generation and our investigation into their transferability across different model architectures shows the breadth and complexity of the threat landscape. The introduction of conformal prediction to Network Intrusion Detection Systems marked a significant innovation in our research. This robust, statistically grounded method enhanced the reliability of our model's predictions by confidently accepting correct predictions and crucially rejecting incorrect ones. The impressive rejection rates of 58.20\% for incorrect predictions in the ISCX dataset and 98.94\% in the ISOT dataset speak volumes about the efficacy of this approach. In future, we plan to explore additional adversarial attack methods that could offer a broader understanding of potential vulnerabilities in Network Intrusion Detection Systems (NIDS). By exposing our models to a broader array of attack vectors, we can further strengthen their resilience and improve their detection capabilities. While conformal prediction has shown promise in enhancing the reliability of NIDS, further research could focus on refining this approach.

\section{Data availability}
The datasets used in this study are essential for validating our proposed methodologies and are publicly available, ensuring transparency and reproducibility of our results. Specifically, we utilized the following datasets: ISOT Botnet Dataset \cite{Saad2011DetectingPB} and ISCX 2014 Botnet Dataset \cite{BiglarBeigi2014EffectiveFS}. The dataset can be downloaded after requesting the owner of the dataset.




 \bibliographystyle{elsarticle-harv} 

\end{document}